\documentstyle[preprint,aps,epsfig,psfig]{revtex}
\begin{document}

      \draft

 \title{Phenomenological Analysis of $B{\rightarrow}PP$ Decays
        with QCD Factorization
 \footnote{Supported in part by National Natural Science Foundation of
        China.}}
\vspace{2cm}
 \author{Dongsheng Du,$^{1,2}$
         Haijun Gong, $^{2}  $
         Junfeng Sun, $^{2}  $
         Deshan Yang, $^{2}  $
         and Guohuai Zhu $^{2}  $ \\
 {\small\em 1. CCAST (World Laboratory),
               P.O.Box 8730,
               Beijing 100080, China} \\
 {\small\em 2. Institute of High Energy Physics,
               Chinese Academy of Sciences,} \\
 {\small\em    P.O.Box 918(4),
               Beijing 100039, China}
  \footnote{Mailing address}
  \footnote{Email:duds@mail.ihep.ac.cn, gonghj@mail.ihep.ac.cn,
  sunjf@mail.ihep.ac.cn,
                  yangds@mail.ihep.ac.cn,
                  zhugh@mail.ihep.ac.cn}}
 \date{\today}

\maketitle

\begin{abstract}
\tighten
\indent

In this paper, we study nonleptonic charmless $B$ decays to two
light pseudoscalar mesons within the frame of QCD factorization,
including the contributions from the chirally enhanced power
corrections and weak annihilation. Predictions for the CP-averaged
branching ratios and CP-violating asymmetries are given. Within
the reasonable range of the parameters, we find that our
predictions for the branching ratios of $B \rightarrow PP$ are
consistent with the present experimental data. But because of the
logarithmic divergences at the endpoints in the hard spectator
scatterings and weak annihilation, there are still large
uncertainties in these predictions.
\end{abstract}

\vspace{1.5cm}

{\bf PACS numbers 13.25.Hw 12.38.Bx}

\newpage

      \narrowtext 
      \tighten

\section{Introduction}
In the standard model (SM), CP violation and quark mixing are
closely related to each other. The unitarity of the
Cabibbo-Kobayashi-Maskawa (CKM) matrix implies various relations
among its elements. The most commonly studied is the ``unitarity
triangle'':
$V_{ud}V_{ub}^{\ast}+V_{cd}V_{cb}^{\ast}+V_{td}V_{tb}^{\ast}=0$.
The study of $B$ meson decays is mainly to make enough independent
measurements for the sides and angles ($\alpha$, $\beta$ and
$\gamma$) of this unitarity triangle. For example, in principle,
$\sin 2\beta$ can be determined by measurements of the time
dependent CP asymmetries of the decay $\overline{B}^0_d\to J/\Psi
K_S$ and other related modes with an uncertainty less than $1\%$.
On the other hand, we can also extract or give a constraint for
the angle $\alpha$ from the decay $B \rightarrow \pi \pi$, and the
angle $\gamma$ from $B \rightarrow \pi \pi$ and $\pi K$. Compared
with determination of $\sin 2\beta$, the extractions of $\alpha$
and $\gamma$ from exclusively non-leptonic charmless $B$ decays
suffer from significant theoretical uncertainties for the strong
interactions.

   Experimentally, many $B$ experiment projects have been running (CLEO,
BaBar, Belle, etc.), or will run in forthcoming years (BTeV, CERN
LHCb, DESY HeraB, etc.). Several collaborations have reported
their latest results recently \cite{cleo,babar,belle}, and more
$B$ decay channels will be measured with great precision soon.
Among these decay channels, quite a few are nonleptonic charmless
two-body modes, such as $B \to \pi K$, $\pi \pi$, $K
\eta^{\prime}$, etc. With the accumulation of the experimental
data, the theorists are urged to gain a deep insight into rare
hadronic $B$ meson decays, and to reduce the theoretical errors in
determining the CKM parameters from the experimental data.

Theoretically, the nonleptonic charmless two-body $B$ decays have
been widely discussed and carefully studied in the framework of
the effective Hamiltonian and factorization hypothesis. The
effective Hamiltonian is obtained by operator product expansion
(OPE) and the renormalization group (RG) method, and generally
expressed as the product of the Wilson coefficients and the
effective operators, in forms of the four-quarks and magnetic
moment operators. The Wilson coefficients can be calculated
reliably by the perturbation theory. In the SM, they have been
evaluated to next-to-leading order\cite{buras1}. Thus, the main
task for us is to compute the hadronic matrix elements of the
effective operators. However, the complexity of QCD dynamics does
not allow us to compute them directly from first principles.
Generally, we resort to the factorization approximation\cite{BSW},
ignoring the soft interactions between the ejected meson and
recoiling system. Under this approximation, the hadronic matrix
element of the effective operator can be parameterized into the
product of the meson decay constant and the meson-meson transition
form factor. A comprehensive analysis for nonleptonic charmless
two-body $B$ decays based on the naive factorization and/or
generalized factorization was done in the past decade, and
achieved great success.

The factorization approximation does hold in the limit that the
soft interactions in the initial and final states can be ignored.
It seems that the argument of color-transparency can give
reasonable support to the above limit\cite{bjorken}. Because the
$b$ quark is heavy, the quarks from $b$ quark decay move so fast
that a pair of quark and antiquark in a small color-singlet object
decouples from the soft interactions. But for the case of naive
factorization the shortcomings are obvious. First, the
renormalization scheme and scale dependence in the hadronic matrix
elements of the effective operators are apparently missed. Then
the full decay amplitude predicted by the Bauer-Stech-Wirbel (BSW)
model remains dependent on the renormalization scheme and scale,
which are mainly from the Wilson coefficients. In recent years,
many researchers have generalized the naive factorization scheme
and made remarkable progress, such as the scheme and scale
independent effective Wilson coefficients, effective color number,
which is introduced to compensate the ``non-factorizable''
contributions, etc.\cite{GF}. Furthermore, some progress in
nonperturbative methods, such as lattice QCD, QCD sum rules etc.
\cite{lattice,pball,ruckle}, allow us to compute many
non-perturbative parameters in $B$ decays, such as the meson decay
constants and meson-meson transition form factors. Every
improvement allows us to have a closer look at the $B$ nonleptonic
decays. In \cite{ali,chy}, the authors gave a comprehensive and
detailed analysis for the nonleptonic two-body charmless $B$
decays within the frame of the generalized factorization.

Two year ago, Beneke, Buchalla, Neubert, and Sachrajda (BBNS) gave
a QCD factorization formula in the heavy quark limit for the
decays $B \to \pi \pi$\cite{beneke}. They pointed out that the
radiative corrections from hard gluon exchange can be
systematically calculated by use of the perturbative QCD method in
the heavy quark limit, i.e., neglecting the power contributions of
$\Lambda_{QCD}/m_b$. This factorization formula can be justified
in the case that the ejected meson from the $b$ quark decay is a
light meson or an onium, no matter whether the other recoiling
meson which absorbs the spectator quark in the $B$ meson is light
or heavy. But for the case that the ejected meson is in an
extremely asymmetric configuration, such as a $D$ meson, this
factorization formula does not hold. The contributions from hard
scattering with the spectator quark in the $B$ meson are also
involved in their formula. This kind of contribution cannot be
contained in the naive factorization, but it appears at the order
of $\alpha_s$. So the naive factorization can be recovered if one
neglects the radiative corrections and suppressed power
$\Lambda_{QCD}/m_b$ contributions in the QCD factorization. The
non-factorizable contributions in the naive factorization can be
calculated perturbatively, so we do not need a phenomenological
parameter $N_c^{eff}$ to compensate the non-factorizable effects.
Moreover, because the $b$ quark mass is not very large in reality,
the potential power corrections, in particular, the chirally
enhanced corrections, will play an important role in $B\to PP$
\cite{osaka,ours2,0104110}. So the twist-3 light-cone wave
function of the light pseudoscalar must be taken into account.
Unfortunately, QCD factorization breaks down at twist-3 level.
There is a logarithmic divergence in the hard spectator scattering
at the endpoint of the twist-3 light-cone distribution amplitude
(LCDA).

Phenomenologically, the QCD factorization (BBNS approach) has been
applied to study many $B$ meson decay modes, such as $B \to
D^{(*)} \pi^-$\cite{chay,nucl}, $\pi \pi$, $\pi
K$\cite{our,ymz,osaka,0104110}, and other interesting
channels\cite{ymz1,hxg,chay1,chy1}. In particular, in a recent
paper \cite{0104110}, Beneke {\it et al.} gave a detailed
phenomenological analysis for $B \to \pi \pi$ and $\pi K$ within
the frame of QCD factorization, in which the contributions from
weak annihilation are also taken into account. In principle, weak
annihilation is power suppressed within the frame of QCD
factorization. Nevertheless, as emphasized in Ref.\cite{lihn},
these contributions may be numerically important for the decays
$B\to PP$. An explicit calculation shows that there are very
significant endpoint divergences in weak annihilation. This
indicates that the weak annihilation may be dominated by the soft
contributions. In Ref.\cite{0104110}, the authors give a
phenomenological treatment of the endpoint divergences in both the
hard spectator scattering and the weak annihilations. Such a
treatment is really a {\it model-dependent} procedure. However, it
may be helpful to estimate the importance of annihilation. The
numerical results show that weak annihilation indeed gives evident
corrections to the branching ratios of the decays $B\to \pi \pi$
and $\pi K$ in Ref.\cite{0104110}, and also causes very large
uncertainties in the prediction. Assuming the universality of the
endpoint divergence, the authors of Ref.\cite{0104110} also gave a
global fit for the CKM parameters by use of the branching ratios
of $B\to \pi \pi$ and $\pi K$ in the frame of QCD factorization,
and derived a new constraint for the ($\bar{\rho}$, $\bar{\eta}$)
complex plane. The new constraint is slightly different from that
derived by the other standard fit, but they are compatible with
each other.

In this work, we will apply QCD factorization to all modes of $B$
decay into two light pseudoscalar mesons, and give a careful and
detailed analysis. For the decay modes $B\to \pi \pi$ and $\pi K$,
our work is actually a re-check of Beneke {\it et
al.}\cite{0104110}, and we find that our results are consistent
with theirs. The $B \to P \eta^{(\prime)}$ has also been studied
by M. Z. Yang and Y. D. Yang within the frame of QCD
factorization\cite{ymz1}. The differences between their work and
ours are that the other type twist-3 light-cone distribution
amplitude $\phi_{\sigma}$ is taken into account, which guarantees
gauge invariance of our results; and the contributions from weak
annihilation are also considered in our work. In addition, the
digluon mechanism for the production of $\eta^{(\prime)}$ is taken
into account in Ref.\cite{ymz1}. Generally, it is believed that
the anomalous coupling of $\eta^{\prime}g^{\ast}g^{\ast}$ can
account for the large branching ratio of $B \to K \eta^{\prime}$.
However, whether the form factor $\eta^{(\prime)} g^{\ast}
g^{\ast}$ is pertubatively calculable or not is still an open
question. In particular,  at the endpoint of the light-cone
distribution amplitudes (LCDAs), where the virtualities  of the
two gluons are both soft, the validity of the suppression from the
LCDAs is questionable. Therefore, in our work, digluon mechanism
is not considered although it can enhance the branching ratio of
$B\to K \eta^{\prime}$. For $B \to K K$, our work is totally new;
the study of these decay modes is expected to give a constraint to
the endpoint divergence in weak annihilation.

This paper is organized as follows. In Sect. II, we give a brief
review of the effective Hamiltonian and QCD factorization
including the chirally enhanced power corrections and weak
annihilations. For a quantitative analysis of $B \to PP$, the
necessary input parameters are discussed in Sect. III. In Sect.
IV, the numerical results for the CP-averaged branching ratios and
CP asymmetries for $B \to PP$ are given and Sect. V is devoted to
the conclusions.

\section{Theoretical frame for $B$ rare decays}
\subsection{The effective Hamiltonian}

      $B$ decays involve three characteristic scales which are strongly
      ordered: $m_W \gg m_b \gg \Lambda_{QCD}$. How to separate or
      factorize these three scales is the most essential question in
      studying $B$ hadronic decays.

        With the operator product expansion method,
       the relevant $\vert\Delta B\vert=1$ effective Hamiltonian
       is given by \cite{buras1}

     \begin{eqnarray}
     {\cal{H}}_{eff}&=& \frac{G_F}{\sqrt{2}}
     \Bigg[ \sum_{q=u,c} v_q \Bigg( C_1(\mu) Q^q_1(\mu)+ C_2(\mu)Q^q_2(\mu)
     + \sum_{k=3}^{10} C_k(\mu)Q_k(\mu)  \Bigg)  \nonumber \\
      && - v_t \Bigg (
     C_{7\gamma}(\mu)Q_{7\gamma}(\mu)+C_{8G}(\mu)Q_{8G}(\mu)
      \Bigg )\Bigg ]+h.c.,
     \end{eqnarray}
      where
      $v_q=V_{qb}V_{qd}^{*}$(for $b\to d$ transition) or
      $v_q=V_{qb}V_{qs}^{*}$(for $b\to s$ transition)
      and $C_i(\mu)$ are the Wilson coefficients which have been evaluated to
      next-to-leading order with
      the perturbation theory and renormalization group method.

       In the Eq.(1), the four-quark operators $Q_i$ are given by
      \begin{equation}
      \begin{array}{l}
      \begin{array}{ll}
      Q^u_1= ( \bar{u}_{\alpha} b_{\alpha} )_{V-A}
               ( \bar{q}_{\beta} u_{\beta} )_{V-A}&
      Q^c_1= ( \bar{c}_{\alpha} b_{\alpha} )_{V-A}
               ( \bar{q}_{\beta} c_{\beta} )_{V-A}\\
      Q^u_2= ( \bar{u}_{\alpha} b_{\beta} )_{V-A}
               ( \bar{q}_{\beta} u_{\alpha} )_{V-A}&
      Q^c_2= ( \bar{c}_{\alpha} b_{\beta} )_{V-A}
               ( \bar{q}_{\beta} c_{\alpha} )_{V-A}\\
      Q_3= (\bar{q}_{\alpha} b_{\alpha} )_{V-A}
            \sum\limits_{q'}
           ( \bar{q}^{'}_{\beta} q^{'}_{\beta} )_{V-A}&
      Q_4= (\bar{q}_{\beta} b_{\alpha} )_{V-A}
            \sum\limits_{q'}
           ( \bar{q}^{'}_{\alpha} q^{'}_{\beta} )_{V-A}\\
      Q_5= (\bar{q}_{\alpha} b_{\alpha} )_{V-A}
            \sum\limits_{q'}
            ( \bar{q}^{'}_{\beta} q^{'}_{\beta} )_{V+A}&
      Q_6= (\bar{q}_{\beta} b_{\alpha} )_{V-A}
            \sum\limits_{q'}
           ( \bar{q}^{'}_{\alpha} q^{'}_{\beta} )_{V+A}\\
      Q_7= \frac{3}{2} (\bar{q}_{\alpha} b_{\alpha} )_{V-A}
            \sum\limits_{q'} e_{q'}
           ( \bar{q}^{'}_{\beta} q^{'}_{\beta} )_{V+A}&
      Q_8=\frac{3}{2}  (\bar{q}_{\beta} b_{\alpha} )_{V-A}
         \sum\limits_{q'} e_{q'}
          ( \bar{q}^{'}_{\alpha} q^{'}_{\beta} )_{V+A}\\
      Q_9= \frac{3}{2} (\bar{q}_{\alpha} b_{\alpha} )_{V-A}
            \sum\limits_{q'} e_{q'}
          ( \bar{q}^{'}_{\beta} q^{'}_{\beta} )_{V-A}&
      Q_{10}=\frac{3}{2}  (\bar{q}_{\beta} b_{\alpha} )_{V-A}
            \sum\limits_{q'} e_{q'}
           ( \bar{q}^{'}_{\alpha} q^{'}_{\beta})_{V-A}\\
      \end{array} \\

      \end{array} ,
      \end{equation}
      and
      \begin{equation}
      Q_{7\gamma}=\frac{e}{8\pi^2} m_b \bar{q}_{\alpha} \sigma^{\mu\nu}
      (1+\gamma_5) b_{\alpha} F_{\mu\nu}, ~~
      Q_{8G}=\frac{g}{8\pi^2} m_b \bar{q}_{\alpha} \sigma^{\mu\nu}
      t^{a}_{\alpha \beta} (1+\gamma_5) b_{\beta} G^a_{\mu\nu}, ~~(q=d~
      {\rm or} ~s).
      \end{equation}
      Here $Q^q_1$ and $Q^q_2$ are the tree operators, $Q_3-Q_6$ the QCD
      penguin operators, $Q_7-Q_{10}$ the electroweak penguin
      operators, and $Q_{7\gamma}$ and $Q_{8G}$ the magnetic-penguin
      operators.

      In this effective Hamiltonian
      for $B$ decays, the contributions from the large virtual momenta of the loop
      corrections from the scale $\mu={\cal O}(m_b)$ to $m_W$ are
      given by the Wilson coefficients, and the low energy contributions
      are fully incorporated into the matrix elements of the
      operators\cite{buras1}. So the derivation of the effective
      Hamiltonian can be called ``the first step factorization''.

         Several years ago, the perturbative corrections to the Wilson
      coefficients in the SM were evaluated to next-to-leading order with
      the renormalization group method. We list the numerical results
      in the naive dimensional regularization (NDR) scheme and at the
      three scales $m_b/2$, $m_b$ and $2m_b$ in Table \ref{tab1}.


\subsection{QCD factorization in the heavy quark limit}
After ``first step factorization'', the
decay amplitude for $B \rightarrow P_1 P_2$ can be written as
      \begin{equation}
      {\cal A} (B \rightarrow P_1 P_2) = \sum \limits_{i} v_i C_i(\mu)
      \langle P_1 P_2 \vert Q_i(\mu) \vert B \rangle ~.
      \end{equation}
The central task now is to reliably evaluate the transition matrix
element $\langle P_1 P_2 \vert Q_i(\mu) \vert B \rangle$. Note
that the matrix element contains two scales: ${\cal O}(m_b)$ and
$\Lambda_{QCD}$. It is hoped that $\langle P_1 P_2 \vert Q_i(\mu)
\vert B \rangle$ can be separated into the short-distance
contributions relating to large scale ${\cal O}(m_b)$ and the
long-distance contributions relating to the fundamental scale of
QCD--$\Lambda_{QCD}$. Then the short-distance contributions, which
are perturbatively calculable, should recover the scheme and scale
dependence of the hadronic matrix elements, and the long-distance
contributions can be parameterized into some universal
non-perturbative parameters.

In Ref.\cite{beneke}, Beneke, Buchalla, Neubert and Sachrajda show
that this idea of factorization can be realized, at least at
one-loop order, in the heavy quark limit. They argue that the
emitted meson $P_1$ carries large energy and momentum (about
$m_B/2$) and therefore can be described by leading-twist
light-cone distribution amplitudes. It is natural to imagine that
the soft gluons would decouple from the emitted $P_1$ at leading
order of $\Lambda_{QCD}/m_b$ since the $q \bar q$ pair in $P_1$
form a small-size color dipole. However, the factorization
requires that not only soft divergences, but also collinear
divergences, should be cancelled. Fortunately, explicit
calculations at one-loop order show that, in the heavy quark
limit, all infrared divergences vanish after summing over the four
vertex correction diagrams [Figs. 1(a)-1(d)]. As for the recoiled
meson $P_2$, because the spectator quark in the $B$ meson is
transferred to it as a soft parton, Beneke {\it et al.} believe
that, instead of light-cone distribution amplitudes, one should
use nonperturbative form factors $F^{BP_2}$ to describe it.
\footnote{Light-cone sum rules also justify this nonperturbative
property for the $B$ to light form factors. It is different from
the conclusion of Keum {\it et al.}\cite{lihn}. In
Ref.\cite{lihn}, the authors claim that the form factor
$F^{B\pi}(0)$ is perturbatively calculable in terms of the
distribution amplitudes when the Sudakov $k_T$ resummation and
threshold resummation are taken into account. However, a recent
detailed examination of such point of view by Descotes-Genon and
Sachrajda shows that it is not justified\cite{Sachrajda}. They
conclude that, the Sudakov form factors are not sufficient to
suppress the contribution that comes from the nonperturbative
region of large impact parameters, and therefore the form factor
$F^{B\pi}(0)$ is uncalculable in the standard hard-scattering
approach.} It should be noted that when the spectator quark
interacts with one of the quarks in the emitted meson by a hard
gluon exchange, the recoiling meson can be described by
leading-twist distribution amplitudes. These hard spectator
diagrams [Figs. 1(g),1(h)] can also be accounted for at the
leading power of $\Lambda_{QCD}/m_b$. In summary, their
factorization formula can be explicitly expressed as
\begin{eqnarray}
\label{qcdf}
\langle P_1 P_2 \vert Q_i \vert B \rangle &=&
F^{B \rightarrow P_2}(0) \int \limits_0^1 dx T^I_i(x) \Phi_{P_1}(x)
+\int \limits_0^1 d\xi dx dy T_i^{II}(\xi,x,y) \Phi_B(\xi) \Phi_{P_1}(x)
\Phi_{P_2}(y) \nonumber \\
&=&\langle P_1 P_2 \vert J_1 \otimes J_2 \vert B \rangle
\cdot [1+ \sum r_n \alpha_s^n +{\cal O}(\Lambda_{QCD}/m_b)].
\end{eqnarray}
We call this factorization formalism as QCD factorization or the
BBNS approach. In Eq.(\ref{qcdf}), $\Phi_B(\xi)$ and
$\Phi_{P_i}(x)$ ($i=1,2$) are the leading-twist wave functions of
$B$ and the light pseudoscalar mesons respectively, and
$T^{I,II}_i$ denote hard-scattering kernels which are calculable
in perturbation theory. At the order of $\alpha_s$, the hard
kernels $T^{I,II}$ can be depicted by Fig. 1. Figures 1(a)-1(d)
represent vertex corrections, Figs. 1(e) and 1(f) penguin
corrections, and Figs. 1(g) and 1(h) hard spectator scattering.

  Comparing this approach with naive factorization and/or generalized
factorization, there are some interesting characteristics in the
BBNS approach.

(i) At the leading order of $\alpha_s$, it can reproduce the
results of naive factorization; at the higher order of $\alpha_s$,
the renormalization scheme and scale dependence for the hadronic
matrix elements can be recovered from the hard-scattering kernels
$T^{I}_{i}$. The generalized factorization can also obtain the
necessary dependence on scheme and scale. However, this dependence
is based on one-loop calculations of quark-level matrix elements.
According to Buras {\it et al.}, quark-level matrix elements are
accompanied by infrared divergences. To avoid such divergences,
one usually assumes that external quark states are off shell.
Unfortunately, this will introduce gauge dependence, which is also
unphysical. In Ref.\cite{chy}, the authors give a calculation
assuming the external quarks are on shell, and they show that the
infrared singularities can be cancelled by assuming the final
quarks are in the form of hadrons; their calculation is therefore
gauge invariant. Their idea is almost the same as the BBNS
approach. The difference is that, in the BBNS approach, the
ejected meson is represented as its light-cone distribution
amplitude; in Ref.\cite{chy}, the ejected $q\bar{q}$ pair is in
such a configuration that it is a color-singlet object and the
quark shares almost the same momentum as the anti-quark. Because
results in Ref.\cite{chy} are still from a calculation at the
quark level, some information from the distributions of momentum
fraction in hadrons is ignored.

(ii) Generalized factorization considers nonfactorizable
contributions as intractable. Therefore, one may introduce one or
more effective color numbers $N_c^{eff}$ to phenomenologically
represent nonfactorizable contributions. Furthermore, $N_c^{eff}$
is assumed to be universal to maintain predictive power. However,
QCD factorization tells us that the nonfactorizable contribution
is indeed factorizable and therefore calculable in the heavy quark
limit. In consequence, $N_c^{eff}$ is calculable, and process
dependent beyond the leading order of $\alpha_s$.

(iii) An interesting result of QCD factorization is that strong
phases come solely from hard scattering processes and are
therefore calculable in the heavy quark limit. From
Eq.(\ref{qcdf}), it is easy to conclude that the imaginary part of
decay amplitude arises only from hard scattering kernels $T^{I}_i$
because nonperturbative form factors and light-cone distribution
amplitudes are all real. $T^{I}_i$ contains vertex corrections
[Figs. 1 (a)-1(d)] and penguin corrections [Figs. 1(e),1(f)].
Strong phases from penguin corrections are commonly called the
Bander-Silverman-Soni (BSS) mechanism\cite{BSS}, which is the
unique source of strong phases in generalized factorization.
However, the gluon virtuality of Fig. 1(e), which is well defined
in QCD factorization, is ambiguous in generalized factorization
and usually treated as a free parameter. In addition, vertex
corrections will also contribute to strong phases in QCD
factorization, which is missed in the generalized factorization.

(iv) Hard spectator contributions [Figs. 1(g),1(h)], which are
leading power effects in QCD factorization, are missing in ``naive
factorization'' and ``generalized factorization''.

Readers are referred to Refs. \cite{beneke,nucl} for more details.

 \subsection{Chirally enhanced corrections in QCD factorization}
 It is observed that QCD factorization is demonstrated only in the strict
heavy quark limit. This means that any generalization of QCD
factorization to include or partly include power corrections of
$\Lambda_{QCD}/m_b$ should redemonstrate the factorization. There
are a variety of sources which may contribute to power corrections
in $1/m_b$; examples are higher twist distribution amplitudes,
transverse momenta of quarks in the light meson, annihilation
diagrams, etc. Unfortunately, there is no known systematic way to
evaluate these power corrections in general for exclusive decays.
Moreover, factorization might break down when these power
corrections, for instance transverse momenta effects, are
considered. This indicates that one might have to give up the
ambitious plan where all power corrections are, at least in
principle, incorporated into QCD factorization order by order. One
might argue that power corrections in $B$ decays are numerically
unimportant because these corrections are proportional to a small
number $\Lambda_{QCD}/m_b \simeq 1/15$. But it is not true. For
instance, the contributions of operator $Q_6$ to decay amplitudes
would formally vanish in the strict heavy quark limit. However it
is numerically very important in penguin-dominant B rare decays,
such as the interesting channels $B \rightarrow \pi K$, etc. This
is because $Q_6$ is always multiplied by a formally power
suppressed but chirally enhanced factor $r_{\chi}=2
m_{P}^2/(m_b(m_1+m_2)) \sim {\cal O}(1)$, where $m_1$ and $m_2$
are current quark masses. So power suppression might probably fail
at least in this case. Therefore phenomenological applicability of
QCD factorization in B rare decays requires at least a consistent
inclusion of chirally enhanced corrections.

The chirally enhanced corrections arise from twist-3 light-cone
distribution amplitudes, generally called $\phi_p(x)$ and
$\phi_{\sigma}(x)$. For light pseudoscalar mesons, they are
defined as \cite{t3}
\begin{eqnarray}
\langle P(p') \vert {\bar q}(y) {\it i} \gamma_5 q(x) \vert 0 \rangle
&=& f_P \mu_P \int_0^1 {\it du~e}^{i(up' \cdot y + {\bar u}p' \cdot x)}
\phi_p(u), \\
\langle P(p') \vert {\bar q}(y) \sigma_{\mu \nu} \gamma_5 q(x) \vert 0
\rangle &=& if_P \mu_P (p^{\prime}_{\mu} z_{\nu} - p^{\prime}_{\nu}
z_{\mu} ) \int_0^1 {\it du~e}^{i(up' \cdot y + \bar u p' \cdot x)}
\frac{\phi_{\sigma}(u)}{6},
\end{eqnarray}
where $\mu_{p}=\frac{m^2_{p}}{m_1+m_2}$, $z=y-x$, and $m_1$ and
$m_2$ are the corresponding current quark masses. So, when
chirally enhanced corrections are concerned, the final light
mesons should be described by leading twist and twist-3
distribution amplitudes \cite{beneke1}:
\begin{eqnarray}
\langle P(p') \vert {\bar q_{\alpha}}(y) q_{\delta}(x) \vert 0 \rangle
&=&\frac{if_P}{4} \int_0^1{\it du~e}^{i(up' \cdot y + {\bar u}p' \cdot
x)} \nonumber \\
&\times& \left \{ \slash{\hskip -2.5mm}p^{\prime} \gamma_5 \phi(u)
-\mu_P \gamma_5 \left ( \phi_p(u)-\sigma_{\mu \nu}p^{\prime \mu}
z^{\nu} \frac{\phi_{\sigma}(u)}{6} \right ) \right \}_{\delta \alpha}.
\end{eqnarray}
Thus it is crucial to show that factorization really holds when
considering twist-3 distribution amplitudes. The most difficult
part is to demonstrate the infrared finiteness of the hard
scattering kernels $T^{I}_i$. For more technical details of this
proof, readers are referred to the literature
\cite{ours2,0104110}.

   With the effective Hamiltonian and QCD factorization formula, $B \to P_1 P_2$ decay amplitudes
in QCD factorization can be written as
     \begin{equation}
     A(B\to P_1 P_2)=\frac{G_F}{\sqrt{2}}
     \sum \limits_{p=u,c} \sum \limits_{i=1,10} v_p a^p_i
     \langle P_1 P_2 \vert Q_i \vert B \rangle_F,
     \end{equation}
 where $v_p$ is the CKM factor, $\langle P_1 P_2 \vert Q_i \vert $B$
\rangle_F$ is the factorized matrix element and is the same as the
definition of the BSW Largrangian\cite{BSW}. The explicit
expressions for the decay amplitudes of $B\to P_1 P_2$ are listed
in the Appendixes of \cite{ali,chy}.

  For illustration, we give the explicit expressions of $a_i^p$ ($i=1$ to
$10$) for $B \to \pi\pi$ (using symmetric LCDAs of the pion). It
is easy to generalize these formulas to the case that the final
states are other light pseudoscalars. Furthermore, we take only
some of QED corrections into account in our final formula, in
particular the QED penguin insertions. Now $a_i^p$ for $B \to \pi
\pi$ in NDR $\gamma_5$ scheme is listed as follows:
      \begin{eqnarray}
      a_1^u&=&C_1+\frac{C_2}{N} + \frac{\alpha_s}{4\pi} \frac{C_F}{N} C_2 F, \\
      a_2^u&=&C_2+\frac{C_1}{N} + \frac{\alpha_s}{4\pi} \frac{C_F}{N} C_1 F,\\
      a_3&=&C_3+\frac{C_4}{N} + \frac{\alpha_s}{4\pi} \frac{C_F}{N} C_4 F, \\
      a_4^p&=&C_4+\frac{C_3}{N} + \frac{\alpha_s}{4\pi} \frac{C_F}{N} C_3 F
      \nonumber \\
      & &- \frac{\alpha_s}{4\pi} \frac{C_F}{N} \left \{
      C_1 (\frac{4}{3}\log\frac{\mu}{m_b}+G(s_p)-\frac{2}{3})+
(C_3-\frac{C_9}{2})(\frac{8}{3}\log\frac{\mu}{m_b}+G(0)+G(1)-\frac{4}{3})
      \right. \nonumber \\
      & & +\sum_{q=u,d,s,c,b}
      (C_4+C_6+\frac{3}{2}{\rm e_q}C_8+\frac{3}{2}{\rm e_q}C_{10}) \left.
      (\frac{4}{3}\log\frac{\mu}{m_b}+G(s_q))+G_8 C_{8G} \right \}, \\
      a_5&=&C_5+\frac{C_6}{N}+\frac{\alpha_s}{4\pi}\frac{C_F}{N} C_6(-F-12),\\
      a_6^p&=&C_6+\frac{C_5}{N} -\frac{\alpha_s}{4\pi} \frac{C_F}{N}6C_5
      \nonumber \\
      & &-\frac{\alpha_s}{4\pi} \frac{C_F}{N} \left \{
      C_1 ((1+\frac{2}{3}A_{\sigma})\log\frac{\mu}{m_b}-\frac{1}{2}-
      \frac{1}{3}A_{\sigma}+G^{\prime}(s_p)+G^{\sigma}(s_p))
      \right. \nonumber \\
      & &+
      \sum_{q=d,b}(C_3-\frac{C_9}{2})
      ((1+\frac{2}{3}A_{\sigma})\log\frac{\mu}{m_b}-\frac{1}{2}-
      \frac{1}{3}A_{\sigma}+G^{\prime}(s_q)+G^{\sigma}(s_q)) \nonumber \\
      & &+\sum_{q=u,d,s,c,b}
      (C_4+C_6+\frac{3}{2}{\rm e_q}C_8+\frac{3}{2}{\rm e_q}C_{10})
      \left( (1+\frac{2}{3}A_{\sigma})\log\frac{\mu}{m_b}
      +G^{\prime}(s_q)+G^{\sigma}(s_q) \right) \nonumber \\
      & & \left. +(\frac{3}{2}+A_{\sigma})C_{8G} \right \}, \\
      a_7&=&C_7+\frac{C_8}{N}+\frac{\alpha_s}{4\pi}\frac{C_F}{N} C_8(-F-12), \\
      a_8^p&=&C_8+\frac{C_7}{N} -\frac{\alpha_s}{4\pi} \frac{C_F}{N}6C_7
      \nonumber \\
      & &-\frac{\alpha_{em}}{9\pi} \left \{
      (C_2+\frac{C_1}{N})
      ((1+\frac{2}{3}A_{\sigma})\log\frac{\mu}{m_b}-\frac{1}{2}-
      \frac{1}{3}A_{\sigma}+G^{\prime}(s_p)+G^{\sigma}(s_p))
      \right. \nonumber \\
      & &+
      (C_4+\frac{C_3}{N}) \sum_{q=d,b} \frac{3}{2}{\rm e_q}
      ((1+\frac{2}{3}A_{\sigma})\log\frac{\mu}{m_b}-\frac{1}{2}-
      \frac{1}{3}A_{\sigma}+G^{\prime}(s_q)+G^{\sigma}(s_q)) \nonumber \\
      & &+(C_3+\frac{C_4}{N}+C_5+\frac{C_6}{N}) \sum_{q=u,d,s,c,b}
      \frac{3}{2}{\rm e_q}
      \left( (1+\frac{2}{3}A_{\sigma})\log\frac{\mu}{m_b}
      +G^{\prime}(s_q)+G^{\sigma}(s_q) \right) \nonumber \\
      & & \left.
      +(\frac{3}{4}+\frac{1}{2}A_{\sigma})C_{7\gamma} \right \}, \\
      a_9&=&C_9+\frac{C_{10}}{N}+\frac{\alpha_s}{4\pi} \frac{C_F}{N} C_{10} F, \\
      a_{10}^{p}&=&C_{10}+\frac{C_9}{N}+
      \frac{\alpha_s}{4\pi} \frac{C_F}{N}C_{9} F
      - \frac{\alpha_{em}}{9\pi} \left \{
      (C_2+\frac{C_1}{N}) (\frac{4}{3}\log\frac{\mu}{m_b}+G(s_p)-\frac{2}{3})
      \right. \nonumber \\
      & &+(C_4+\frac{C_3}{N})\sum_{q=d,b} \frac{3}{2}{\rm e_q}
      (\frac{4}{3}\log\frac{\mu}{m_b}+G(s_q)-\frac{2}{3}) \nonumber \\
      & & +(C_3+\frac{C_4}{N}+C_5+\frac{C_6}{N}) \sum_{q=u,d,s,c,b} \left.
      \frac{3}{2}{\rm e_q}
      (\frac{4}{3}\log\frac{\mu}{m_b}+G(s_q))+\frac{1}{2}G_8 C_{7\gamma}
      \right \}.
      \end{eqnarray}
      Here $N=3$ is the color number,
      $C_F=(N^2-1)/2N$ is the color factor,
      $s_q=m_q^2/m_b^2$, and we define the other symbols
      in the above expressions as
      \begin{eqnarray}
      &&F=-12 \ln \frac{\mu}{m_b} -18+f^{I}+f^{II}, \\
      &&f^{I}=\int \limits_{0}^{1}~ dx~g(x)\phi(x),
      ~G_8=\int \limits_{0}^{1}~ dx~G_8(x) \phi(x), \\
      &&G(s)=\int \limits_{0}^{1}~ dx~G(s,x) \phi(x), \\
      &&G^{\prime}(s)=\int \limits_{0}^{1}~ dx~G^{\prime}(s,x) \phi_p(x),\\
      &&G^{\sigma}(s)=\int \limits_{0}^{1}~ dx~G^{\sigma}(s,x)
      \frac{\phi_{\sigma}(x)}{6(1-x)},~~~~~
      A_{\sigma}=\int \limits_{0}^{1}~ dx~ \frac{\phi_{\sigma}(x)}{6(1-x)},
      \end{eqnarray}
      where $\phi(x)$ [$\phi_p(x)$, $\phi_{\sigma}(x)$] is leading twist
(twist-3)
      LCDA of the ejected pion, and the hard-scattering functions are
      \begin{eqnarray}
      &&g(x)=3 \frac{1-2x}{1-x} \ln x - 3 i \pi, ~~G_8(x)=\frac{2}{1-x}, \\
      &&G(s,x)=-4 \int \limits_{0}^{1}~ du~u(1-u) \ln (s-u(1-u)(1-x)-i
      \epsilon), \\
      &&G^{\prime}(s,x)=-3 \int \limits_{0}^{1}~ du~u(1-u) \ln (s-u(1-u)(1-x)-i
      \epsilon), \\
      &&G^{\sigma}(s,x)=-2 \int \limits_{0}^{1}~ du~u(1-u) \ln (s-u(1-u)(1-x)-i
      \epsilon) \nonumber \\
      &&~~~~~~~~~~~~~~~
      + \int \limits_{0}^{1}~ du~ \frac{u^2(1-u)^2(1-x)}{s-u(1-u)(1-x)-i \epsilon}.
      \end{eqnarray}
      The contributions from the hard spectator scattering [Figs.
1(g), 1(h)] are reduced to the factor $f^{II}$:
       \begin{equation}
      f^{II}=\frac{4 \pi^2}{N}
      \frac {f_{\pi}f_B}{F^{B\to \pi}_{+}(0) m_B^2}
      \int \limits_{0}^{1}~ d\xi~ \frac{\Phi_B(\xi)}{\xi}
      \int \limits_{0}^{1}~ dx~ \frac{\phi(x)}{x} \int \limits_{0}^{1}~
      dy~ \left [ \frac{\phi(y)}{1-y}+\frac{2 \mu_{\pi}}{M_B}
      \frac{\phi_{\sigma}(y)}{6(1-y)^2} \right ].
      \end{equation}

      There is a divergent integral in $f^{II}$.  When we do
numerical calculation in this work, we will simply parameterize
this divergence as what was done by Beneke {\it et
al.}\cite{osaka}:
\begin{equation}
X_H=\int dy/y = \ln (m_b/\Lambda_{QCD})+\varrho_H e^{i\phi_H},
\end{equation}
where $\phi_H$ is an arbitrary phase, $0^{\circ}\leq\phi_H\leq
360^{\circ}$, and $\varrho_H$ is varied from $0$ to $3$
(realistic) or $6$ (conservative). \footnote{This parameterization
is a little bit different from that in the recent
paper\cite{0104110} by Beneke {\it et al.}, but the variation of
these two parameterization are almost the same. In addition, we
pointed out in \cite{ours2} that $f^{II}$ might be convergent if
the transverse momentum $k_T$ of the parton is taken into account
and the Sudakov resummation is invoked. In Ref.\cite{weizt}, the
authors find that value of $f^{II}$ is really in the range of the
above parameterization when the Sudakov suppression is taken into
account. In recent literature\cite{threshold} of the perturbative
QCD (PQCD) method, the authors point out that the threshold
resummation can also help to suppress the endpoint singularities,
especially for the twist-3 level. However, this treatment
contradicts the claim of Descotes-Genon and Sachrajda. They
conclude in Ref.\cite{Sachrajda} that it is impossible to make
reliable predictions for power corrections to the amplitudes for
exclusive two-body $B$ decays which have end-point singularities,
even if the Sudakove effects are taken into account. We think that
this point of view needs further careful investigation.}

       We illustrate numerically the scale dependence of $a_i^p$ in Table
\ref{tab2}. Here we use the asymptotic form of the LCDAs of the
light pseudoscalar meson, which are
\begin{eqnarray}
\phi(x)&=&6x(1-x), \\
\phi_p(x)&=&1, \\
\phi_{\sigma}(x)&=&6x(1-x),
\end{eqnarray}
and we set $f^{II}=0$ in computation because of its uncertainty.

    From the expressions and numerical results of the coefficients $a_i$,
some general observations are listed below.

(1) $a_i$ ($i=1-5,7,9,10$) and $a_{6,8}r_{\chi}$ are RG invariant
so they guarantee a
 full decay amplitude independent of the renormalization scale.
 We have shown that the coefficients $a_{i}$
 ($i=1-5,7,9,10$) are renormalization scale independent at the order of
${\alpha}_{s}$ with asymptotic LCDAs in \cite{ours2},
\[
\frac{{\bf d}a_{i}}{{\bf d}{\ln}{\mu}}=0 ~~~~~(i{\neq}6,8),
\]
However, $a_6$ and $a_8$ must be scale dependent because the
factorized matrix elements multiplied by $a_6$ and $a_8$ are
scale-dependent. But,
\[
\frac{{\bf d}(a_{i}r_{\chi})}{{\bf d}{\ln}{\mu}}=0~~~~~~~~(i=6,8).
\]
This point can also be seen roughly from numerical results in
Table \ref{tab2}. It should be noted that the imaginary part of
$a_i$ is at the order of $\alpha_s$, and its remaining scale
dependence must be cancelled by the radiative corrections from
higher order of $\alpha_s$.

(2) The coefficients $a_i$ are gauge invariant. At the leading
power of $\Lambda_{QCD}/m_b$, the light meson is represented as
its leading twist light-cone wave function (LCWF). In the
calculation, the Dirac structure of the leading twist LCWF
guarantees that the light meson can be taken as a pair of
collinear on-shell massless quark and antiquark. Therefore, the
on-shell condition guarantees the gauge invariance of $a_i$
($i\neq 6,8$). In Ref.\cite{0104110}, the authors have shown that
at the twist-3 level the light meson can be also taken as a pair
of on-shell massless quark and anti-quark when the asymptotic form
of twist-3 LCDAs are taken. Then, $a_{6,8}$ are also independent
of the gauge choice.

(3)The quantities of $a_{1}$, $a_{2}$ receive a special attention,
because they are related to tree operators $Q_{1,2}$, so they are
numerically large. QCD penguins effects are incorporated to the
coefficients $a_{3-6}$, in which $a_{3,5}$ are smaller than
$a_{4,6}$, and $a_{3}+a_{5}$ is ${\cal O}(10^{-4})$. $a_{4,6}$
have the same order as $a_{2}$, so they will play a dominant role
when the effects of tree operators are CKM-suppressed, for example
in the $b{\to}s$ transition. The electroweak penguin coefficients
$a_{7}$-$a_{10}$ are even smaller. $a_{9}$ is the largest one
among them, and numerically comparable with the QCD penguin
coefficients $a_{3,5}$. Thus, $a_9$ may play an important role in
some decay modes, such as $B \to \pi^0 K$.

(4) The imaginary part of $a_i$ arises from the radiative
corrections. Unlike in the generalized factorization, the
imaginary part of $a_i$ comes not only from the penguin
corrections (BSS mechanism), but also from the vertex corrections.
However, it is suppressed by $\alpha_s$, and generally numerically
small. Thus, the strong phases evaluated from the BBNS approach
are generally small. But in some special cases the situation will
be different. For example, the imaginary parts of $a_{2,4,6}$,
especially for $a_{2}$, are larger than those of the other
coefficients, so, when their contributions are dominant in some
decay modes, strong interaction phases might be large, which would
result in a large CP violation.

 \subsection{Contributions of annihilation amplitudes}

Annihilation contributions appear in almost all charmless decay
modes $B \to PP$. In some cases, they may be important in spite of
the power suppression. As emphasized in the PQCD
method\cite{lihn,lucd}, weak annihilation can give a large
imaginary part to the decay amplitudes. Then, within the framework
of the PQCD method, large CP violations are expected. In this
work, we will follow Beneke {\it et al.}\cite{0104110}, writing
the annihilation contributions in terms of convolutions of
``hard-scattering'' kernels with LCDAs by ignoring the soft
endpoint divergences. This kind of treatment is obviously not
self-consistent; however, it can help us to give an estimation of
the annihilation. The corresponding diagrams of weak annihilation
are depicted in Fig. 2. Here, we will follow the convention in
Ref.\cite{0104110}:
 \begin{equation}
 {\cal A}^{ann}\Big(B{\rightarrow}P_{1}P_{2}\Big)
 {\propto}f_{B}f_{P_{1}}f_{P_{2}}{\sum}v_{i}\;r\;b_{i}
 \label{eq:annihilation-2}
 \end{equation}

The decay amplitudes of weak annihilation for $B\to PP$ are listed
in the Appendix.

 Similar to the treatment for the endpoint divergence in hard
 spectator scattering, we also parameterize the endpoint
 divergence $X_A$ in the weak annihilation as follows:
 \begin{equation}
X_A=\int dy/y = \ln (m_b/\Lambda_{QCD})+\varrho_A e^{i\phi_A},
\end{equation}
where $\phi_A$ is an arbitrary phase, $0^{\circ}\leq\phi_A\leq
360^{\circ}$, and $\varrho_A$ is varied from $0$ to $3$
(realistic) or $6$ (conservative). It should be noted that there
is no correlation between $X_A$ and $X_H$ for their different
origins.

We will apply the above formulae to revisit rare $B$ decays in
what follows.


 \section{Input Parameters}
 \label{sec:parameter}
 The decay amplitude for $B{\to}PP$ can be expressed by various
parameters, such as the CKM matrix elements, form factors, Wilson
coefficients $C_{i}({\mu})$, LCDAs, and so on. The values of these
parameters will affect our predictions for CP-averaged branching
ratios and CP-violating asymmetries. Now we will specify them for
use in calculation.

 \subsection{CKM matrix elements}
 \label{sec:CKM}
 The CKM matrix in the Wolfenstein parameterization reads
 \begin{equation}
 V_{CKM}=\left(\begin{array}{ccc}
     1-{\lambda}^{2}/2
  &    {\lambda}
  &   A{\lambda}^{3}({\rho}-i{\eta})      \\
      -{\lambda}
  &  1-{\lambda}^{2}/2
  &   A{\lambda}^{2}                          \\
      A{\lambda}^{3}(1-{\rho}-i{\eta})
  &  -A{\lambda}^{2}
  &  1  \end{array}\right) +{\cal O}(\lambda^4).
 \label{eq:ckm-1}
 \end{equation}
 Here we take the Wolfenstein parameters from the fit of Ciuchini et al.
\cite{roma}:
               $A=0.819 {\pm}0.040$,
       $\lambda=0.2237{\pm}0.0033$,
 $\overline{\rho}=\rho (1-\lambda^2/2)=0.224{\pm}0.038$,
 $\overline{\eta}=\eta (1-\lambda^2/2) =0.317{\pm}0.040$ and
        $\gamma=(54.8{\pm}6.2)^{\circ}$.
 Correspondingly, we have
   $\rho=0.230{\pm}0.039$, $\eta=0.325{\pm}0.039$, and
   $\sqrt{\rho^2+\eta^2}=0.398{\pm}0.040$.
 If not stated otherwise, we shall use the central values as the default
 values.

 \subsection{Form factors and decay constants}
 \label{sec:formfactor}
  Decay constants and heavy-to-light form factors are defined by following
current matrix elements:

\begin{equation}
\langle P (q)\vert \bar{q}_1 \gamma_{\mu} \gamma_5 q_2 \vert 0 \rangle =
-i f_P q_{\mu},
\end{equation}
\begin{eqnarray}
\langle P (q) \vert \bar {q} \gamma_{\mu} b \vert B \rangle &=&
\left[ (p+q)_{\mu}-\frac{m_B^2-m_P^2}{(p-q)^2}(p-q)_{\mu} \right] F_1((p-q)^2)\nonumber \\
&&+\frac{m_B^2-m_P^2}{(p-q)^2}(p-q)_{\mu} F_0((p-q^2).
\end{eqnarray}

 For the decay constants we use $f_{\pi}=131$ MeV, $f_{K}=160$ MeV, and $f_{B}=180\pm 40$ MeV.
 As to the decay constants related to the ${\eta}$ and ${\eta}^{\prime}$,
  we shall take the convention in \cite{ali,chy}:
\[
 \langle 0 \vert \bar{q} \gamma^{\mu} \gamma_{5} q \vert
 \eta^{(\prime)}(p) \rangle =i f_{\eta^{(\prime)}}^q p^{\mu},
\]
 and
 \[
 \langle 0 \vert \bar{s} \gamma_5 s \vert \eta^{(\prime)}(p) \rangle
 =-i\frac{(f_{{\eta}^{({\prime})}}^{s}-f_{{\eta}^{({\prime})}}^{u})
    m_{{\eta}^{({\prime})}}^{2}}{2m_{s}}
,\]

\[
\langle 0 \vert \bar{u} \gamma_5 u \vert \eta^{(\prime)}(p) \rangle
=\langle 0 \vert \bar{d} \gamma_5 d \vert \eta^{(\prime)}(p) \rangle
=\frac{f_{\eta^{(\prime})}^{u}}{f_{\eta^{(\prime})}^{s}}
\langle 0 \vert \bar{s} \gamma_5 s \vert \eta^{(\prime)}(p) \rangle.
 \]
 Here we shall not consider the charm quark content in $\eta^{(\prime)}$.
 The relations between ${\eta}$-${\eta}^{\prime}$ mixing and
${\eta}_{8}$-${\eta}_{0}$ [the $SU(3)$ octet and singlet] are
 \begin{equation}
 \vert \eta {\rangle}= \cos  \theta_8 \vert  \eta_{8} \rangle
                       -\sin \theta_0 \vert \eta_0 \rangle,
  \hspace*{10mm}
 \vert \eta^{\prime} \rangle=
       \sin \theta_8 \vert \eta_8 \rangle
      +\cos \theta_0 \vert \eta_0 \rangle,
 \end{equation}
 \begin{equation}
 f^{u}_{\eta}=\frac{f_8}{\sqrt{6}} \cos \theta_8
             -\frac{f_0}{\sqrt{3}} \sin \theta_0,
 \hspace*{10mm}
 f^{u}_{{\eta}^{\prime}}=\frac{f_8}{\sqrt{6}} \sin \theta_8
                        +\frac{f_0}{\sqrt{3}} \cos \theta_0,
 \end{equation}
 \begin{equation}
 f^{s}_{\eta}=-2\frac{f_8}{\sqrt{6}} \cos \theta_8
              - \frac{f_0}{\sqrt{3}} \sin \theta_0,
 \hspace*{10mm}
 f^{s}_{{\eta}^{\prime}}=-2\frac{f_8}{\sqrt{6}} \sin \theta_8
                          +\frac{f_0}{\sqrt{3}} \cos \theta_0,
 \end{equation}
 Here $f_8$, $f_0$ are decay constants of $\eta_8$, $\eta_0$
respectively, $\theta_8$, $\theta_0$ are the mixing angles. Their
values are\cite{feldman}
         $f_8 \approx 1.28 f_{\pi} \approx 168$ MeV,
         $f_0 \approx1.20 f_{\pi} \approx157$ MeV,
  $\theta_8 \approx-22.2^{\circ}$, and
  $\theta_0 \approx-9.1^{\circ}$.

 We take
$F^{B{\pi}}_{0}=0.28{\pm}0.05$, which is obtained by the
light-cone sum rule\cite{ruckle}. For $B{\to}K$,
${\eta}^{({\prime})}$ transitions, the following relations are
applied to give their corresponding values:
 \begin{equation}
 \frac{f_{\pi}}{f_{K}}{\approx}\frac{F_{0}^{B{\pi}}(0)}{F_{0}^{BK}(0)},
 \hspace*{5mm}
 F_{0,1}^{B{\eta}}=F_{0,1}^{B{\pi}}\left(
    \frac{{\cos}{\theta}_{8}}{\sqrt{6}}
   -\frac{{\sin}{\theta}_{0}}{\sqrt{3}}\right),
 \hspace*{5mm}
 F_{0,1}^{B{\eta}^{\prime}}=F_{0,1}^{B{\pi}}\left(
    \frac{{\sin}{\theta}_{8}}{\sqrt{6}}
   +\frac{{\cos}{\theta}_{0}}{\sqrt{3}}\right).
 \end{equation}

 \subsection{Quark masses}
 \label{sec:quark-mass}
 There are two types of quark mass in our analysis. One type is the pole
mass which appears in the loop integration. Accompanied by the
$\cal{O}(\alpha_s)$ corrections to the hadronic matrix elements,
the pole masses contribute to coefficients $a_{i}$ through the
functions $G(s,x)$, $G^{\prime}(s,x)$ and $G^{\sigma}(s,x)$, or
$G(s)$, $G^{\prime}(s)$ and $G^{\sigma}(s)$. Here we fix them as
 \[
  m_{u}=m_{d}=m_{s}=0,~~
  m_{c}=1.45 \hbox{GeV},~~
  m_{b}=4.6 \hbox{GeV}.
 \]
The other type quark mass appears in the hadronic matrix elements
and the chirally enhanced factor $r_{\chi}$ through the equations
of motion. They are renormalization scale dependent. And their
values have been summarized in \cite{0010175}. Here we shall use
the 2000 Particle Data Group (PDG2000) data for discussion:
 \[
 {\overline{m}}_{b}({\overline{m}}_{b})=4.0\sim 4.4 \hbox{GeV},
 ~~
 {\overline{m}}_{s}(2{\rm GeV})=75 \sim 170 \hbox{MeV},~~
 \]
 \[
 {\overline{m}}_{d}(2{\rm GeV})=3\sim 9\hbox{MeV},    ~~
 {\overline{m}}_{u}(2{\rm GeV})=1\sim 5\hbox{MeV}.
 \]
Here, we take the central values of the $b$quark and $s$ quark as
default values. Because the current masses of light quarks are
difficult to fix, we would like to take
\[
r_{\eta^{(\prime)}}(1-\frac{f_{\eta^{(\prime)}}^u}
{f_{\eta^{(\prime)}}^s})= r_{\pi} = r_{K}=r_\chi.
\]

We think that it is a good approximation in our calculation.
Using the renormalization group equation, we can get the
corresponding running $r_\chi$ at the scale ${\mu}={\cal O}(m_{b})$:

\[
r_\chi(m_b/2)=0.85, ~~~~~ r_\chi(m_b)=1.14,~~~~~ r_\chi(2m_b)=1.42.
\]

 \section{Numerical Analysis}
 \label{sec:numerical-analysis}

 \subsection{Branching ratios and comparison with data}
 \label{sec:tables-analysis}

In the $B$ meson rest frame, we have two-body decay width
\begin{equation}
{\Gamma}(B{\to}P_{1}P_{2})=\frac{1}{8{\pi}}
\frac{{\vert}p{\vert}}{m_{B}^{2}}
{\vert}{\cal A}(B{\to}P_{1}P_{2}){\vert}^{2},
\label{eq:width}
\end{equation}
where
\[ {\vert}p{\vert}=\frac{\sqrt{\Big[m_{B}^{2}-(m_{P_{1}}+m_{P_{2}})^{2}
   \Big]\Big[m_{B}^{2}-(m_{P_{1}}-m_{P_{2}})^{2}\Big]}}{2m_{B}} .
\]
The corresponding branching ratio is
\begin{equation}
{\rm BR}(B{\to}P_{1}P_{2})=\frac{{\Gamma}(B{\to}P_{1}P_{2})}
{{\Gamma}_{tot}}, \label{eq:partial-branching-ratios}
\end{equation}
where ${\tau}=1/{{\Gamma}_{tot}}$. In our calculation, we take
${\tau}=1.548{\times}10^{-12}$ sec for  $B^0$, and
${\tau}=1.653{\times}10^{-12}$ sec for $B^{\pm}$. In addition,
because we work in the heavy quark limit, we take the mass of
light meson as zero in the computation of phase space; then $\vert
p \vert=m_B/2$.

We show experimental data and theoretical predictions for the
CP-averaged branching ratios for $B{\rightarrow}PP$ in Table
\ref{tab3} and Table \ref{tab4}, respectively. In Table
\ref{tab4}, we give the averages of the branching ratios of
$B{\rightarrow}PP$ and their CP-conjugate modes with the default
values of the parameters at three different renormalization scales
$\mu=m_b/2$, $m_b$, $2m_b$. From Table \ref{tab3} and Table
\ref{tab4} some remarks are in order.

 (1) For the decay modes $B\to \pi \pi$ and $B \to \pi K$, our
predictions are consistent with those by Beneke {\it et
al.}\cite{0104110}. In these modes, with the default values of the
parameters, we find that the CP-averaged branching ratios of
$B^{\pm} \to \pi^0 \pi^-$ ($\simeq$ pure tree) and $B^{\pm}\to K^0
\pi^{\pm}$ (pure penguin) are in good agreement with the
experimental measurements. Therefore these two decay modes can be
considered as good probes of QCD factorization for $B$ meson
decays to two light mesons\cite{lp01}. For the other modes $B\to
\pi \pi$ and $B \to \pi K$, the predictions of the CP-averaged
branching ratios obtained by using the default values of the
parameters seem not very good compared with the experimental
measurements. In particular, the theoretical prediction for the
branching ratio of $B^0 \to \pi^+ \pi^-$ is larger than the
experimental data. The same situation also occurs in generalized
factorization\cite{ali,chy}. For the other decays modes $B\to \pi
K$, our predictions seem smaller than the measurements. But taking
the uncertainties of the input parameters into account, such as
the CKM matrix elements (especially the angle $\gamma$ and $\vert
V_{ub}\vert$), the endpoint divergences in hard spectator
scattering and weak annihilations, form factors, etc., it is
possible that the theoretical predictions from the BBNS approach
can be in accord with the experimental results. In
Refs.\cite{0110093,lp01}, the author gave a global fit to the
CP-averaged branching ratios of $B\to \pi \pi$, $\pi K$ within the
framework of the QCD factorization. With the best fit values, the
branching ratios of $B\to \pi \pi$ and $\pi K$ are in very good
agreement with the experimental measurements very well. [Although,
in the best fit, value of $\gamma$ is slightly too large (about
$90^{\circ}$) and $\vert V_{ub}\vert$ is comparably small, they
are still compatible with the standard fit using semileptonic
decays, $K-\overline{K}$ mixing and $B-\overline{B}$ mixing.]

 (2) We note that experimental data for CP-averaged branching ratios
for $B{\to}K{\eta}^{\prime}$ are about four times larger than our
theoretical predictions with default values of the parameters. The
results under the generalized factorization (GF) framework are
similar \cite{ali,chy}. In Refs.\cite{ali,chy}, the charm quark
content in $\eta^{\prime}$ is considered; however, this mechanism
cannot give a good explanation for experimental data yet. The
unexpectedly large data have triggered considerable theoretical
interest in understanding the mechanism of ${\eta}^{\prime}$
production, and have been widely discussed in the
literature\cite{yyd}. So far, it is generally assumed that digluon
fusion mechanism could enhance ${\eta}^{\prime}$ production, but
the shape of form factor of the vertex
$g^{\ast}g^{\ast}\eta^{\prime}$ brings a very large uncertainty to
predictions of the contribution from digluon mechanism \cite{ff}.
Recently, M.Z. Yang and Y.D. Yang gave a calculation for $B \to
\eta^{(\prime)} P$ within the framework of the BBNS approach
\cite{ymz1}, in which the digluon mechanism is considered. In that
paper, the authors compute the vertex of $g^{\ast}g^{\ast}
\eta^{\prime}$ in perturbative QCD, and find the branching ratios
of $B \to K \eta^{\prime}$ are really enhanced and in agreement
with the experimental measurement. But, as mentioned in our
Introduction, we think that the consistency of their perturbative
calculation is questionable due to the endpoint behavior.

  (3) For the decays $B^{\pm}\to \pi^{\pm} \eta^{(\prime)}$, $K^{\pm}
\eta$, the CP-averaged branching ratios predicted by the BBNS
approach are at the order of $10^{-6}$. We expect that these decay
modes can be observed soon at the BaBar and Belle experiments
\footnote{Recently, Belle and BaBar collaborations report their
measurements of the branching ratio of the decay $B^{\pm} \to
\pi^{\pm} \eta^{\prime}$ which is consistent with zero. The $90\%$
CL upper limit is ${\rm BR} (B^{\pm}\to \pi^{\pm} \eta^{\prime})<
7\times 10^{-6}$ (at Belle) or  ${\rm BR}(B^{\pm}\to \pi^{\pm}
\eta^{\prime})< 12\times 10^{-6}$(at BaBar).}.

 (4) For the decays $B^{0}\to \pi^0 \pi^0$, $\pi^0 \eta^{(\prime)}$,
$\eta^{(\prime)} \eta^{(\prime)}$, our predictions show that their
CP-averaged branching ratios are very small and about ${\cal
O}(10^{-7})\sim {\cal O}(10^{-8})$. If these predictions are
reliable, it is very hard to observe these decay modes at the
BaBar and Belle experiments.

(5) In the BBNS approach, the predictions for the branching ratios
of $B\to K K$ are highly suppressed, especially for $B^0 \to K^+
K^-$, which is purely from the weak annihilation contributions. In
Ref.\cite{lihn2}, the authors evaluate the decays $B\to K K$ by
use of the PQCD method, and also found that their branching ratios
are very small. These decay modes can be a good probe for
long-distance final state interaction (FSI). On the other hand, as
we show in the Appendix, if the endpoint divergence $X_A$ is
universal, the measurement for the branching ratio of $B^0\to K^+
K^-$ may give a constraint for the range of $X_A$.

In Table \ref{tab4}, for comparison, we also give the estimations
for the CP-averaged branching ratios of $B\to PP$ using the naive
factorization (NF) approximation. Again, we list the predictions
by the BBNS approach with and without considering the weak
annihilation also. From these results, some general observations
are given as follows.

Obviously, predictions on branching ratios for $B{\to}PP$ using
the BBNS approach are less scale dependent compared with those in
the NF framework, especially in the $b{\to}s$ transition, such as
$B{\to}K{\pi}$, $K{\eta}^{({\prime})}$.  Noting that $a_{i}$ are
calculated at one-loop level, if the corrections from high order
of ${\alpha}_{s}$ were taken into account, the scale dependence of
theoretical predictions might be further reduced.

 Generally, the contributions from weak annihilation are
highly power suppressed. However, due to the endpoint divergences,
they may give observable contributions in our results. In $B\to
\pi K$, we can see that the weak annihilation can give about
$5\%\sim 10\%$ enhancement. Note that the results of Table
\ref{tab4} are computed by the default parameters. If we varied
the parameters for the endpoint divergences in weak annihilations,
it would give much larger uncertainties to our predictions. This
will be shown in the next subsection in detail.

 \subsection{Uncertainty of predictions}
 \label{sec:figure-analysis}
 In Fig. 3, based on the BBNS approach, we show the dependence of
branching ratios for $B \to \pi \pi$, $\pi K$ and $K
\eta^{\prime}$ on the weak phase ${\gamma}$, scanning all ranges
of the variation of parameters, including the CKM matrix elements,
form factors, endpoint divergence ${\int}^{1}_{0} dx /x$, and so
on. Note that weak annihilation effects do not contribute to
$B^{\pm} \to \pi^{\pm}\pi^0$. We can see that our theoretical
predictions give very wide dot shades for the branching ratios
within the range of parameters. So it is possible that the
predictions for CP-averaged branching ratios of the decays $B
\rightarrow \pi \pi$, $\pi K$ are simultaneously consistent with
the experimental data by use of the same input parameters. The
results of the global fit in Refs.\cite{0104110,lp01,0110093} show
that, with the best fit values of the input parameters and within
the framework of QCD factorization, the predictions for the
branching ratios of $B\to \pi \pi$ and $\pi K$ can fit the
experimental data simultaneously with good $\chi^2/n_{dof}$.

  In Fig. 4, we show the dependence of the CP-averaged branching
ratios of $B^0 \to \pi^0 K^0$, $B^{\pm} \to K^{\pm} \pi^0$ and
$B^0 \to \pi^+ \pi^-$ on the CKM matrix elements, form factors,
and endpoint divergence in hard spectator scattering and weak
annihilations. From Fig. 4, some remarks are given in order.

 For the decays $B^0 \to \pi^0 K^0$ and $B^{\pm} \to K^{\pm} \pi^0$,
the CP-averaged branching ratios weakly depend on $\vert
V_{ub}/V_{cb}\vert$ because that the decays $B\to \pi K$ are
dominated by penguin diagrams and $\vert v_u/v_c\vert=\vert
(V_{us}^{\ast}V_{ub})/(V_{cs}^{\ast}V_{cb})\vert$ is highly
suppressed by ${\cal O}(\lambda^2)$ in the $b\to s$ transition.
For the case of $B^0 \to \pi^+ \pi^-$, although it is a
tree-dominant ($b\to u$) process, the penguin pollution plays an
important role. For illustration, the decay amplitude of
$\bar{B}^0 \to \pi^+\pi^-$ reads
\begin{eqnarray}
{\cal A}(\bar{B}^0 \to \pi^+ \pi^-)&=&-i \frac{G_F}{\sqrt{2}} f_{\pi}
F^{B\pi}(0)m_B^2 \Big \{v_u\Big[a_1+a_4^u+a_{10}^u+\Big (a_6^u+a_8^u\Big)
r_{\chi}\Big ]\nonumber \\
&&+v_c\Big[a_4^c+a_{10}^c+\Big (a_6^c+a_8^c\Big)
r_{\chi}\Big ]\Big\}.
\end{eqnarray}
Note that $\vert v_u\vert=\vert V_{ud}^{\ast} V_{ub}\vert$ is at
the same order as $\vert v_c\vert=\vert V_{cd}^{\ast}V_{cb}\vert$
and $\vert (a_4+a_6 r_{\chi})/a_1\vert \sim 0.1$, so the branching
ratio of $\bar{B}^0\to \pi^+\pi^-$ is sensitive to both $\vert
V_{ub}/V_{cb}\vert$ and $\gamma = \arg (V_{ub}^{\ast})$. (In
contrast, the decay amplitude of $B^{\pm} \to \pi^0 \pi^{\pm}$
suffers less pollution from the penguin contribution, which is
proportional to the small electroweak penguin coefficient
$a_9+a_{10}-a_7+a_8 r_{\chi}$.) Therefore, the variation of $\vert
V_{ub}/V_{cb}\vert$ gives large uncertainty to the prediction for
the branching ratio of $B^0\to \pi^+\pi^-$.

All the three modes are dominated by the form factor
$F^{B\pi}(0)$. Varying $F^{B\pi}(0)$ from $0.23$ to $0.33$, it
gives about $\pm 35\%$ uncertainty to the branching ratios. By
using the ratios of the branching ratios, the uncertainties caused
by the form factor can be largely reduced. This will be shown
later.

  The endpoint divergence in the hard spectator scattering does not
bring large uncertainty to our predictions. The reason is that the
endpoint divergence contains only a single logarithmic term and
its coefficient is suppressed by $\alpha_s$ and the color number
$N$.

 Obviously, the endpoint divergence in weak annihilation causes large
uncertainty, in particular, to the decays $B \to \pi K$. In
Ref.\cite{0104110}, the authors also showed this point. The reason
is that the endpoint divergence in weak annihilation is dominated
by a product of two logarithmic terms. Within the range of the
parameters $\varrho$ and $\phi$, our predictions by the BBNS
approach are in agreement with the experimental measurements.

   Therefore, we find that weak annihilation is the main source
of the uncertainties in the BBNS approach. To obtain a more
precise prediction, further analysis is needed of the weak
annihilations. On the other hand, there are other possible power
suppressed corrections in the decays $B\to PP$. Some researchers
pointed out that the elastic and inelastic long-distance FSI can
help us obtain small branching ratio for $\bar{B}_d^0\to \pi^+
\pi^-$ at small $\gamma$\cite{fsi,xing}. However, with the elastic
FSI, in order to get satisfying results, one must introduce a
large strong phase difference which is unbelievable in the heavy
quark limit. In particular, in Ref.\cite{xing}, the author takes
the inelastic rescattering $B \to D \bar{D} \to \pi \pi$ into
account and finds that the small branching ratio of
$\bar{B}^0_d\to\pi \pi$ can be obtained at small $\gamma$ without
fine-tuning the input parameters. Furthermore, in
Ref.\cite{charming}, Isola {\it et al.} give detailed research on
the nonperturbative charming penguin effects on $B\to \pi K$ and
the predictions are in better agreement with the experimental
data. However, all treatments to these power corrections are
model-dependent. Up to now, there is no systematic way to deal
with these complicated nonperturbative interactions. Maybe the
precise measurement of the CP asymmetries for the decays $B\to PP$
is a chance to distinguish these models.

  Here we emphasize that the decay amplitude of $B^0 \to K^{-}K^{+}$
comes completely from the effects of weak annihilations under the
BBNS approach, because contributions of weak annihilations are
supposed to be power suppressed, so the CP-averaged branching
ratio for $\overline{B}^{0}(B^0)\to K^{+}K^{-}$ is small in Fig.
3. At the same time, we can also see that uncertainties of
theoretical predictions for these decay modes are very large (the
branching ratios range from $10^{-8}$ to $10^{-7}$). As mentioned
above, we expect that we can extract some useful information about
the endpoint divergence $X_A$ from $B^0\to K^+ K^-$. However, the
present experimental upper limit is much greater than our
predictions within the range of the parameters. If this decay mode
can be measured precisely in the future, then we will be able to
obtain some useful information on weak annihilations and FSI.
Consequently, the endpoint divergence $X_A$ can be determined
effectively.

 \subsection{Ratios of the branching ratios}
 As mentioned above, the decay rates for $B{\to}PP$ depend on a number of
parameters that cannot be predicted with the BBNS approach itself.
Thus, these parameters result in ambiguity of the predictions for
the CP-averaged branching ratios. In order to reduce uncertainties
from the renormalization scale and weak annihilation effects,
etc., as much as possible, the ratios of CP-averaged branching
ratios are good quantities. Furthermore, many methods have been
suggested to give bounds on ${\gamma}$ using decay modes of
$B{\rightarrow}{\pi}{\pi}$, ${\pi} K$, in which the ratios of
branching ratios play important roles. We hope that our
calculation can give some constraints on $\gamma$. In Fig. 5, we
show some ratios of CP-averaged branching ratios versus the weak
phase ${\gamma}$ (note that we use PDG2000 data for
${\tau}_{B^{+}}/{\tau}_{B^{0}}=1.062$), where the slashed-line
bands correspond to experimental data listed in Table \ref{tab3}
with one standard deviation ($1\sigma$).

Obviously, using the ratios of branching ratios, and comparing
with Fig. 3, the width of dot-shades becomes narrow due to the
reduction of the uncertainty. From Fig. 5(b), the ratio $ {\rm
BR}(B^0 \to \pi^{\mp} K^{\pm})/(2{\rm BR}(B^0 \to \pi^0 K^0))$ can
be made consistent with the experimental data at small $\gamma$.
However, the ratio $\tau_{B^+}/\tau_{B^0}
 [{\rm BR}(B^0\to \pi^+ \pi^-)/
      (2{\rm BR}(B^0 \to \pi^{\pm} \pi^0))]$
tends to be consistent with the experimental data at large
$\gamma$ in Fig. 5(e). Considering the correlation between these
quantities, a scientific choice is to give a global fit for these
branching ratios and/or the ratios of the branching ratios. In
Ref.\cite{0104110}, the authors give a global fit for
$(\bar{\rho},\bar{\eta})$ from $B$ rare hadronic decays with the
QCD factorization. The result favors large $\gamma$ (about
$90^{\circ}$) and/or small $\vert
V_{ub}\vert$\cite{0104110,lp01,0110093}.

 There are many discussions on ratios of branching ratios, for
instance in \cite{ali,chy}, in order to determine coefficients
$a_{i}$, and extract the CKM matrix parameters, etc. In
particular, the decay modes $B{\to}K{\pi}$ are widely discussed to
get information on ${\gamma}$; readers interested in these topics
are referred to, for instance, \cite{neubert,fleisher}. From Figs.
5(a) and 5(b), our theoretical results show that
\begin{equation}
 \frac{2{\rm BR}(B^{\pm}{\to}{\pi}^{0}K^{\pm})}
      { {\rm BR}(B^{\pm}{\to}{\pi}^{\pm}K^{0})}
       {\approx}
 \frac{ {\rm BR}(B^{0}{\to}{\pi}^{\mp}K^{\pm})}
      {2{\rm BR}(B^{0}{\to}{\pi}^{0}  K^{0}  )}
\label{eq:figure5a-b}
\end{equation}
This approximate relation can also be obtained from the isospin
decomposition of the effective Hamiltonian \footnote{In
Ref.\cite{neubert1}, the isospin amplitudes are defined as
$B_{1/2}=\sqrt{2/3} \langle 1/2, \pm 1/2 \vert {\cal H}_{\Delta
I=0} \vert 1/2, \pm 1/2 \rangle$, which arises from QCD penguin
operators, while $A_{1/2}=\pm \sqrt{2/3} \langle 1/2, \pm 1/2
\vert {\cal H}_{\Delta I=1} \vert 1/2, \pm 1/2 \rangle$,
$A_{3/2}=\sqrt{1/3} \langle 3/2, \pm 1/2 \vert {\cal H}_{\Delta
I=1} \vert 1/2, \pm 1/2 \rangle$, which include contributions from
tree and electroweak operators. Moreover,
${\vert}B_{1/2}{\vert}{\approx}{\cal O}(10^{-3})$,
${\vert}A_{1/2}{\vert}{\approx}{\cal O}(10^{-5})$, and
${\vert}A_{3/2}{\vert}{\approx}{\cal O}(10^{-4})$. The decay
amplitudes can be described as ${\cal
A}(B^{+}{\to}{\pi}^{+}K^{0})=A_{3/2}+A_{1/2}+B_{1/2}$, and
$\sqrt{2}{\cal
A}(B^{+}{\to}{\pi}^{0}K^{+})=2A_{3/2}-A_{1/2}-B_{1/2}$. Neglecting
effects of $A_{3/2}$ and $A_{1/2}$, we can get a good
approximation to Eq.(\ref{eq:figure5a-b}).} \cite{neubert1}. The
experimental data listed in Table \ref{tab3} give
\begin{equation}
\frac{2{\rm BR}({\pi}^{0}K^{\pm})}{{\rm BR}({\pi}^{\pm}K^{0})}
=1.18{\pm}0.34\hbox{(BaBar)}, ~~~~~~~~~ \frac{{\rm
BR}({\pi}^{\mp}K^{\pm})}{2{\rm BR}({\pi}^{0}K^{0})}
=1.02{\pm}0.41\hbox{(BaBar)},
\end{equation}
which are consistent with Eq.(\ref{eq:figure5a-b}).

 \subsection{CP-violating asymmetries}
 \label{cp-vioating}

 CP violation in $B$ meson decays can be either direct or indirect.
 Typically, direct CP-violating asymmetry arises when the tree and
 penguin amplitudes interfere. CP violations in decays of
 charged $B$ meson are purely direct; as these decays are all
 self-tagging, they are the definitive signals of CP violation if seen.
 The direct CP-violating symmetry is defined by
 \begin{equation}
 {\cal A}_{CP}=
  \frac{{\Gamma}(B^{-}{\to}f^{-})-{\Gamma}(B^{+}{\to}f^{+})}
       {{\Gamma}(B^{-}{\to}f^{-})+{\Gamma}(B^{+}{\to}f^{+})}.
 \label{eq:direct-cp}
 \end{equation}

 Because the flavor eigenstates ${\overline{B}}^{0}$ and $B^{0}$ mix due to
 weak interactions, direct and indirect CP violation
 occur simultaneously in neutral $B$ meson decays, and
 time-dependent measurements of CP-violating asymmetries are
 needed.
 \begin{equation}
  {\cal A}_{CP}(t)=
  \frac{{\Gamma}(\overline{B}^{0}(t)\to \bar{f})
       -{\Gamma}({B}^{0}(t){\to} f )}
       {{\Gamma}(\overline{B}^{0}(t){\to}\bar{f})
       +{\Gamma}({B^{0}(t)\to f })}.
 \label{eq:time-integration-cp-one}
 \end{equation}
 For $B^{0}{\to}f$ (but $B^{0}{\not\to}{\bar{f}})$ and
 ${\overline{B}}^{0}{\to}{\bar{f}}$ (but
    ${\overline{B}}^{0}{\not\to}f)$,
 the CP-violating asymmetries are similar to those of charged
 $B$ mesons. For
 ${\overline{B}}^{0}{\rightarrow}(\bar{f}=f, CP
  {\vert}f{\rangle}={\pm}{\vert}f{\rangle}){\leftarrow}B^{0}$,
 i.e., the final states are CP eigenstates, the time-integrated
 CP-violating asymmetries are:
 \begin{equation}
 {\cal A}_{CP}=\frac{1}{1+x^{2}}a_{{\epsilon}^{\prime}}+
 \frac{x}{1+x^{2}}a_{{\epsilon}+{\epsilon}^{\prime}},
 \label{eq:time-integration-cp-two}
 \end{equation}
  \begin{equation}
 a_{{\epsilon}^{\prime}}=
 \frac{1-{\vert}{\lambda}_{f}{\vert}^{2}}
      {1+{\vert}{\lambda}_{f}{\vert}^{2}}, \;
 a_{{\epsilon}+{\epsilon}^{\prime}}=
 \frac{-2\hbox{Im}({\lambda}_{f})}
      {1+{\vert}{\lambda}_{f}{\vert}^{2}}, \;
 {\lambda}_{f}=
 \frac{V_{tb}^{\ast}V_{td}}{V_{tb}V_{td}^{\ast}}
 \frac{{\langle}f{\vert}{\cal H}_{eff}{\vert}
       {\overline{B}}^{0}{\rangle}}
      {{\langle}f{\vert}{\cal H}_{eff}{\vert}B^{0}{\rangle}},
 \label{eq:time-integration-cp}
 \end{equation}
 with the PDG2000 data value $x={\Delta}m/{\Gamma}=0.73{\pm}0.03$ for
 $B^{0}$-${\overline{B}}^{0}$ system. Here
 $a_{{\epsilon}^{\prime}}$ and
 $a_{{\epsilon}+{\epsilon}^{\prime}}$ are due to direct and
 mixing-induced CP violation, respectively.

 Recent experimental measurements of CP-violating asymmetries ${\cal
A}_{CP}$ for $B{\to}{\pi}K$, $K{\eta}^{\prime}$ and
$a_{\epsilon^{\prime}}$, $a_{\epsilon+\epsilon^{\prime}}$ for
$\bar{B}^0_d\to \pi^+\pi^-$ are listed in Table \ref{tab5}. These
results contain very large errors and they are consistent with
zero. Precise measurements of direct CP violation and
time-dependent CP violation are expected in the near future.

Theoretically, CP-violating asymmetries will not be very large in
the BBNS approach in principle, because the strong phases are
suppressed by ${\alpha}_{s}$ or ${\Lambda}_{QCD}/m_{b}$. However,
from Table \ref{tab2} we know that $a_{2,4,6}$ have large
imaginary parts. When they are dominant in some decay modes, large
CP violations
 are expected, for instance, in
 ${\overline{B}}^{0}(B^{0}){\to}{\pi}^{0}{\pi}^{0}$,
 ${\pi}^{0}{\eta}^{({\prime})}$, ${\eta}^{({\prime})}{\eta}^{({\prime})}$.

 With the BBNS approach, CP-violating asymmetries depend on several
 variables, such as the form factors, CKM parameters, the scale ${\mu}$, and
 so on. Our investigation indicates that CP-violating asymmetries
 for all $B{\to}PP$ depend very weakly on form factors; the same
 conclusion can be obtained with the GF approach. So we show results
 only with the default values of the form factors in the following
 calculations. As for the dependence of CP-violating asymmetries on
 scale ${\mu}$, there are six decay modes
 ${\overline{B}}^{0}(B^{0}){\to}{\pi}^{0}{\pi}^{0}$,
 ${\pi}^{0}{\eta}^{({\prime})}$, ${\eta}^{({\prime})}{\eta}^{({\prime})}$
 to be singled out because they are strongly ${\mu}$ dependent
 compared with other decay modes. In these six decay modes,
 $a_{2}$ is involved and not CKM-suppressed, and it is sensitive
 to the renormalization scale. This point has been shown in Table \ref{tab2}.
In particular, its imaginary part is larger than the real part,
which might result in large CP violations in these decay modes as
we show in Table \ref{tab6}. We also list CP-violating asymmetries
versus the weak phase ${\gamma}$ in Table \ref{tab7} and Table
\ref{tab8}. The endpoint divergence can cause large theoretical
uncertainties; its influence on CP-violating asymmetries is shown
in Table \ref{tab9}.

 From Table \ref{tab6} to Table \ref{tab9}, we can clearly see that theoretical
 predictions on CP-violating asymmetries are compatible with
 measurements within one standard deviation ($1\sigma$).
Unfortunately, the
 present experimental uncertainties are too large to draw any
 meaningful conclusion. In addition, the weak annihilations
 have great influence on CP-violating asymmetries, so their
 contributions to decay amplitudes must be included in estimation
 of CP violation in $B$ meson decays with the BBNS approach. Moreover,
 the power corrections in order of ${\Lambda}_{QCD}/m_{b}$ which are not
 included in the QCDF master formula (\ref{qcdf}). This needs further
investigation.

 Here we would like to point out that $B {\to} {\pi}{\pi} $ are usually
considered to determine the CKM angles ${\alpha}$ and ${\gamma}$
using CP-violating asymmetry\cite{fleisher1,bigi}. From Table
\ref{tab7} and Table \ref{tab8}, we can see that ${\cal
A}({\pi}^{+}{\pi}^{-})$, especially the mixing-induced
$a_{{\epsilon}+{\epsilon}^{\prime}}({\%})$ term, are sensitive to
the CKM angle ${\gamma}$. This mode is strongly polluted by the
penguins. As to the large penguin pollution, recently, Fleischer
suggested a new approach \cite{fleisher1}. Using U-spin SU(3)
flavour symmetry, $B_{d} {\to}{\pi}^{+}{\pi}^{-}$ and
$B_{s}{\to}K^{+}K^{-}$ are related to each other by interchanging
the $d$ quark and $s$ quark, which possibly allows a simultaneous
determination ${\beta}$ and ${\gamma}$. His analysis gives
${\gamma}=76^{\circ}$ and ${\phi}_{d}=2{\beta}=53^{\circ}$ without
considering the U-spin-breaking effects.

  As we see from Table \ref{tab7} and Table \ref{tab8}, time-integrated CP-violating
asymmetries ${\cal A}_{CP}$ for
${\overline{B}}^{0}{\rightarrow}K_{S}^{0}{\pi}$,
$K_{S}^{0}{\eta}^{({\prime})}$ are completely dominated by
mixing-induced $a_{{\epsilon}+{\epsilon}^{\prime}}$ term. But we
still do not know the ${\eta}^{({\prime})}$ production mechanism.
Hence it is useless to discuss the CP-violating asymmetries ${\cal
A}_{CP}(K_{S}^{0}{\eta}^{({\prime})})$. As to CP-violating
asymmetries for ${\overline{B}}^{0}{\rightarrow}K^{+}K^{-}$, they
do not contain the direct CP-violating $a_{\epsilon}^{\prime}$
term, and arise only from $a_{{\epsilon}+{\epsilon}^{\prime}}$
term. Moreover, they are sensitive to the CKM angle ${\gamma}$.
However, we know that with the BBNS approach the branching ratios
for neutral $B$ meson decays into $K^{+}K^{-}$ have only weak
annihilations contributions, which are scale ${\mu}$ dependent and
contain large theoretical uncertainties due to their soft
nonperturbative effects.

 \section{Conclusions}

(1) In the heavy quark limit, neglecting the effects of order
${\Lambda}_{QCD}/m_{b}$, the contributions that are
non-factorizable under the GF framework can be factorized with the
BBNS approach, and they are perturbatively calculable from first
principles, at least at order of ${\alpha}_{s}$. The BBNS approach
provides decay amplitudes with strong phases which are usually
small, at order of ${\alpha}_{s}$ and/or ${\Lambda}_{QCD}/m_{b}$.

(2) CP-averaged branching ratios of $B{\to}PP$ in the BBNS
approach are less renormalization scale dependent than those under
the NF framework, especially when the chirally enhanced
corrections are taken into account; generally, with the
appropriate parameters, predictions with the BBNS approach are
consistent with present experimental data. However, the
theoretical predictions include very large uncertainties which
arise from various parameters, such as the form factors, CKM
matrix elements and soft endpoint divergence, etc. Our results
show that the theoretical uncertainties from the endpoint
divergence $X_A$ are considerable. Likewise, CP-violating
asymmetries for $B{\to}PP$ in the BBNS approach are also sensitive
to variation of endpoint divergences. In particular, decay
amplitudes for ${\overline{B}}^{0}(B^{0}){\to}K^{+}K^{-}$ arise
completely from weak annihilations with the BBNS approach; we can
obtain very useful information on the effects of FSI and soft weak
annihilations with precise experimental data in the future.

(3) The present experimental data for non-leptonic $B$ meson
decays are not precise enough yet, especially for the measurements
of CP violation. On the other hand, there are still many
uncertainties in the theoretical framework, for instance, the hard
spectator scattering, weak annihilations and other potential power
corrections. Great advances in both experiment and theory in the
near future are expected to determine the CKM angles precisely.

 \section*{Acknowledgements}
 This work is Supported in part by National Natural Science Foundation of
 China. We thank Profs. Zhi-zhong Xing and Mao-Zhi Yang for helpful
discussions.

 \begin{appendix}
 \section*{The annihilation amplitudes for $B{\to}PP$}
 \label{sec:annihilation-amplitudes}
 \begin{eqnarray}
  A^{ann}({\overline{B}}^{0}{\to}{\pi}^{+}{\pi}^{-})&=&
  -i\frac{G_{F}}{\sqrt{2}}f_{B}f^{2}_{\pi}\bigg[v_{u}b_{1}
  +(v_{u}+v_{c})
  \Big(b_{3}+2b_{4}-\frac{1}{2}b_{3}^{ew}+\frac{1}{2}b_{4}^{ew}
  \Big)\bigg],
  \\
  A^{ann}({\overline{B}}^{0}{\to}{\pi}^{0}{\pi}^{0})&=&
  A^{ann}({\overline{B}}^{0}{\to}{\pi}^{+}{\pi}^{-}),
  \\
  A^{ann}(B^{-}{\to}{\pi}^{0}{\pi}^{-})&=&0,
  \\
  A^{ann}({\overline{B}}^{0}{\to}{\pi}^{0}{\overline{K}}^{0})&=&
  i\frac{G_{F}}{2}f_{B}f_{\pi}f_{K}\bigg[(v_{u}+v_{c})
  \Big(b_{3}-\frac{1}{2}b_{3}^{ew}\Big)\bigg],
  \\
  A^{ann}({\overline{B}}^{0}{\to}{\pi}^{+}K^{-})&=&
  -\sqrt{2}A^{ann}({\overline{B}}^{0}{\to}{\pi}^{0}{\overline{K}}^{0}),
  \\
  A^{ann}(B^{-}{\to}{\pi}^{0}K^{-})&=&
  -i\frac{G_{F}}{2}f_{B}f_{\pi}f_{K}\bigg[v_{u}b_{2}
  +(v_{u}+v_{c})\Big(b_{3}+b_{3}^{ew}\Big)\bigg],
  \\
  A^{ann}(B^{-}{\to}{\pi}^{-}{\overline{K}}^{0})&=&
  \sqrt{2}A^{ann}(B^{-}{\to}{\pi}^{0}K^{-}),
  \\
  A^{ann}({\overline{B}}^{0}{\to}{\pi}^{0}{{\eta}^{({\prime})}})&=&
  -i\frac{G_{F}}{2}f_{B}f_{\pi}f_{{\eta}^{({\prime})}}^{u}
  \bigg[v_{u}2b_{1}
  +(v_{u}+v_{c})\Big(-2b_{3}+b_{3}^{ew}+3b_{4}^{ew}\Big)\bigg],
  \\
  A^{ann}(B^{-}{\to}{\pi}^{-}{\eta^{({\prime})}})&=&
  -i\frac{G_{F}}{\sqrt{2}}f_{B}f_{\pi}f_{\eta^{({\prime})}}^{u}
  \bigg[v_{u}2b_{2}
  +(v_{u}+v_{c})\Big(2b_{3}+2b_{3}^{ew}\Big)\bigg],
  \\
  A^{ann}({\overline{B}}^{0}{\to}{\eta}^{({\prime})}{\eta}^{({\prime})})
  &=&
  -i\frac{G_{F}}{\sqrt{2}}f_{B}{f_{{\eta}^{({\prime})}}^{u}}^{2}
  \bigg[v_{u}2b_{1}
  +(v_{u}+v_{c})\Big(2b_{3}+4b_{4}-b_{3}^{ew}+b_{4}^{ew}\Big)\nonumber
  \\
  &&+(v_u+v_c)\Big(\frac{f_{\eta^{(\prime)}}^s}{f_{\eta^{(\prime)}}^u}\Big)^2
  \Big(2b_4-b_4^{ew}\Big)\bigg],
  \\
  A^{ann}({\overline{B}}^{0}{\to}{\eta}{\eta}^{({\prime})})&=&
  -i\frac{G_F}{\sqrt{2}}f_Bf_{\eta}^u f_{\eta^{\prime}}^u
  \bigg[v_{u}2b_{1}
  +(v_{u}+v_{c})\Big(2b_{3}+4b_{4}-b_{3}^{ew}+b_{4}^{ew}\Big)\nonumber
  \\
  &&+(v_u+v_c)\frac{f_{\eta^{\prime}}^s f_{\eta}^s}{f_{\eta^{\prime}}^u f_{\eta}^u}
  \Big(2b_4-b_4^{ew}\Big)\bigg],
  \\
  A^{ann}(B^{-}{\to}K^{-}{\eta}^{({\prime})})&=&
  -i\frac{G_{F}}{\sqrt{2}}f_{B}f_{K}f_{{\eta}^{({\prime})}}^{u}
  \bigg(1+\frac{f_{{\eta}^{({\prime})}}^{s}}
               {f_{{\eta}^{({\prime})}}^{u}}\bigg)
  \bigg[v_{u}b_{2}
  +(v_{u}+v_{c})\Big(b_{3}+b_{3}^{ew}\Big)\bigg],
  \\
  A^{ann}({\overline{B}}^{0}{\to}{\overline{K}}^{0}{\eta}^{({\prime})})
  &=&
  -i\frac{G_{F}}{\sqrt{2}}f_{B}f_{K}f_{{\eta}^{({\prime})}}^{u}
  \bigg(1+\frac{f_{{\eta}^{({\prime})}}^{s}}
               {f_{{\eta}^{({\prime})}}^{u}}\bigg)
  \bigg[(v_{u}+v_{c})\Big(b_{3}-\frac{1}{2}b_{3}^{ew}\Big)\bigg],
  \\
  A^{ann}({\overline{B}}^{0}{\to}{\overline{K}}^{0}K^{0})&=&
  -i\frac{G_{F}}{\sqrt{2}}f_{B}f_{K}^{2}
  \bigg[(v_{u}+v_{c})
  \Big(b_{3}+2b_{4}-\frac{1}{2}b_{3}^{ew}-b_{4}^{ew}\Big)\bigg],
  \\
  A^{ann}(B^{-}{\to}K^{-}K^{0})&=&
  -i\frac{G_{F}}{\sqrt{2}}f_{B}f_{K}^{2}\bigg[v_{u}b_{2}
  +(v_{u}+v_{c})\Big(b_{3}+b_{3}^{ew}\Big)\bigg],
  \\
  A^{ann}({\overline{B}}^{0}{\to}K^{+}K^{-})&=&
  -i\frac{G_{F}}{\sqrt{2}}f_{B}f_{K}^{2}\bigg[v_{u}b_{1}
  +(v_{u}+v_{c})\Big(2b_{4}+\frac{1}{2}b_{4}^{ew}\Big)\bigg],
  \end{eqnarray}
 The annihilation coefficients $(b_{1},b_{2})$, $(b_{3},b_{4})$ and
 $(b_{3}^{ew},b_{4}^{ew})$ correspond to the contributions from tree,
 QCD penguins and electroweak penguins operators, respectively. They are
 related to final state mesons. Using the asymptotic light cone
 distribution amplitudes of the mesons, and assuming $SU(3)$ flavor
 symmetry, they can be expressed as \cite{0104110}
 \begin{eqnarray}
  b_{1}&=&\frac{C_{F}}{N_{c}^{2}}C_{1}A^{i},
  \hspace*{10mm}
  b_{3}=\frac{C_{F}}{N_{c}^{2}}\bigg[C_{3}A^{i}
       +A^{f}\Big(C_{5}+N_{c}C_{6}\Big)\bigg],
   \\
  b_{2}&=&\frac{C_{F}}{N_{c}^{2}}C_{2}A^{i},
  \hspace*{10mm}
  b_{4}=\frac{C_{F}}{N_{c}^{2}}A^{i}\Big(C_{4}+C_{6}\Big),
   \\
  b_{3}^{ew}&=&\frac{C_{F}}{N_{c}^{2}}\bigg[C_{9}A^{i}
             +A^{f}\Big(C_{7}+N_{c}C_{8}\Big)\bigg],
   \\
  b_{4}^{ew}&=&\frac{C_{F}}{N_{c}^{2}}A^{i}\Big(C_{10}+C_{8}\Big),
 \end{eqnarray}

 and
 \begin{eqnarray}
  A^{i}&{\approx}&{\pi}{\alpha}_{s}\bigg[18\Big(X_{A}-4+
        \frac{{\pi}^{2}}{3}\Big)+2r_{\chi}^{2}X_{A}^{2}\bigg]
   \\
  A^{f}&{\approx}&12{\pi}{\alpha}_{s}r_{\chi}\Big(2X_{A}^{2}-X_{A}\Big),
  \end{eqnarray}
  where $X_{A}={\int}_{0}^{1}{\bf dy}/y$ is a logarithmically divergent integral,
  and will be phenomenologically parameterized in the calculation.
 \end{appendix}

 \begin{table}
 \caption{Wilson coefficients in NDR scheme. The input parameters
          in numerical calculation are fixed:
          ${\alpha}_{s}(m_{Z})=0.1185$, ${\alpha}_{em}(m_{W})=1/128$,
          $m_{W}=80.42\text{GeV}$, $m_{Z}=91.188\text{GeV}$,
          $m_{t}=168.2\text{GeV}$, $m_{b}=4.6\text{GeV}$.}
 \label{tab1}
 \begin{tabular}{cccc}
            & ${\mu}=m_{b}/2 $
            & ${\mu}=m_{b}   $
            & ${\mu}=2m_{b}  $      \\ \hline
  $C_{1} $  & 1.136 & 1.080 & 1.044 \\
  $C_{2} $  &-0.283 &-0.181 &-0.105 \\
  $C_{3} $  & 0.021 & 0.014 & 0.009 \\
  $C_{4} $  &-0.050 &-0.035 &-0.024 \\
  $C_{5} $  & 0.010 & 0.009 & 0.007 \\
  $C_{6} $  &-0.063 &-0.041 &-0.026 \\
  $C_{7}/{{\alpha}_{em}} $ &-0.020 &-0.004 & 0.019 \\
  $C_{8}/{{\alpha}_{em}} $ & 0.082 & 0.052 & 0.033 \\
  $C_{9}/{{\alpha}_{em}} $ &-1.339 &-1.263 &-1.201 \\
  $C_{10}/{{\alpha}_{em}}$ & 0.369 & 0.253 & 0.168 \\
  $C_{7{\gamma}}         $ &-0.341 &-0.304 &-0.272 \\
  $C_{8g}                $ &-0.160 &-0.145 &-0.132
 \end{tabular}
 \end{table}

 \begin{table}
 \caption{Numerical values of coefficients $a_{i}$ with
          $m_b=4.6\text{GeV}$, $m_c=1.45\text{GeV}$ and
          $f^{II}=0$.}
 \label{tab2}
 \begin{tabular}{rccc}
   & ${\mu}=m_{b}/2$
   & ${\mu}=m_{b}  $
   & ${\mu}=2m_{b} $ \\ \hline
  $a_{1}^{u}\ \ \ \ $
      &  1.070+0.026i & 1.046+0.013i & 1.027+0.006i \\
  $a_{2}^{u}\ \ \ \ $
      & -0.019-0.106i & 0.023-0.079i & 0.062-0.064i \\ \hline
  $a_{3}(10^{-4})$
      & 93.662+46.585i & 73.837+25.729i & 51.675+14.646i  \\
  $a_{4}^{u}(10^{-4})$
      &-330.78-179.23i &-297.17 -146.84i &-267.83 -125.46i \\
  $a_{4}^{c}(10^{-4})$
      &-403.19-55.903i &-351.48-54.337i &-311.41 -51.226i  \\
  $a_{5}(10^{-4})$
      &-98.679-58.697i &-67.453-30.139i &-41.612 -15.867i  \\
  $a_{6}^{u}(10^{-4})$
      &-556.39-159.84i & -411.61-136.65i & -322.84 -120.02i \\
  $a_{6}^{c}(10^{-4})$
      &-597.2-34.652i & -442.22-42.742i & -347.41 -44.663i \\
  $a_{6}^{u}r_{\chi}(10^{-4})$
      &-474.17-136.22i  & -469.29-155.8 i & -459.82-170.95i \\
  $a_{6}^{c}r_{\chi}(10^{-4})$
      &-508.95-29.531i  & -504.19-48.731i & -494.81-63.614i \\ \hline
  $a_{7}(10^{-4})$
      & 0.458+0.597i & 1.248+0.299i & 2.591+0.157i  \\
  $a_{8}^{u}(10^{-4})$
      & 6.980-0.66i  &  4.255-1.117i & 2.336-1.465i \\
  $a_{8}^{c}(10^{-4})$
      & 6.884-0.365i &  4.076-0.566i & 2.092-0.717i \\
  $a_{8}^{u}r_{\chi}(10^{-4})$
      & 5.949-0.562i &  4.852-1.274i & 3.328-2.087i \\
  $a_{8}^{c}r_{\chi}(10^{-4})$
      & 5.867-0.311i &  4.647-0.645i & 2.980-1.021i \\
  $a_{9}(10^{-4})$
      &-97.902-2.686i &-94.935-1.453i &-91.732-0.801i  \\
  $a_{10}^{u}(10^{-4})$
      & 7.414+9.087i & 3.136+6.136i & -0.956+4.261i  \\
  $a_{10}^{c}(10^{-4})$
      & 7.244+9.377i & 2.817+6.679i &-1.389+4.998i
  \end{tabular}
  \end{table}

 \begin{table}
 \caption{Experimental data of CP-averaged branching ratios for
          $B{\to}{\pi}{\pi}$, ${\pi}K$, $K{\eta}^{\prime}$
          in unit of $10^{-6}$.}
 \label{tab3}
 \begin{tabular}{lccc}
 Decay Modes & CLEO  \cite{cleo}
             & BaBar \cite{babar}
             & Belle \cite{belle} \\ \hline
  ${\overline{B}}^{0}{\to}{\pi}^{+}{\pi}^{-}$ &
  $4.3^{+1.6}_{-1.4}{\pm}0.5 $ &
  $4.1{\pm}1.0{\pm}0.7       $ &
  $5.6^{+2.3}_{-2.0}{\pm}0.4 $ \\
  $B^{\pm}{\to}{\pi}^{\pm}{\pi}^{0}$ &
  $5.6^{+2.6}_{-2.3}{\pm}1.7 $ &
  $5.1^{+2.0}_{-1.8}{\pm}0.8 $ &
  $7.8^{+3.8+0.8}_{-3.2-1.2} $ \\
  ${\overline{B}}^{0}{\to}{\pi}^{\pm}K^{\mp}$ &
  $17.2^{+2.5}_{-2.4}{\pm}1.2  $ &
  $16.7{\pm}1.6{\pm}1.3        $ &
  $19.3^{+3.4+1.5}_{-3.2-0.6}  $ \\
  $B^{\pm}{\to}{\pi}^{0}K^{\pm}$ &
  $11.6^{+3.0+1.4}_{-2.7-1.3}  $ &
  $10.8^{+2.1}_{-1.9}{\pm}1.0  $ &
  $16.3^{+3.5+1.6}_{-3.3-1.8}  $ \\
  $B^{\pm}{\to}{\pi}^{\pm}K^{0}$ &
  $18.2^{+4.6}_{-4.0}{\pm}1.6  $ &
  $18.2^{+3.3}_{-3.0}{\pm}1.7  $ &
  $13.7^{+5.7+1.9}_{-4.8-1.8}  $ \\
  ${\overline{B}}^{0}{\to}{\pi}^{0}{\overline{K}}^{0}$ &
  $14.6^{+5.9+2.4}_{-5.1-3.3}  $ &
  $8.2^{+3.1}_{-2.7}{\pm}1.1   $ &
  $16.0^{+7.2+2.5}_{-5.9-2.7}  $ \\
  $B^{-}{\to}K^{-}{\eta}^{\prime}$ &
  $80^{+10}_{-9}{\pm}7 $  &
  $70{\pm}8{\pm}5      $  &
  $79^{+12}_{-11}{\pm}9$ \\
  ${\overline{B}}^{0}{\to}{\overline{K}}^{0}{\eta}^{\prime}$ &
  $89^{+18}_{-16}{\pm}9$  &
  $42^{+13}_{-11}{\pm}4$  &
  $55^{+19}_{-16}{\pm}8$
  \end{tabular}
  \end{table}

 \begin{table}
 \caption{Numerical predictions for CP-averaged branching ratios (in
          unit of $10^{-6}$) for $B{\to}PP$, in the framework of NF and
          QCDF, where ${{\rm BR}}^{f}$ and ${{\rm BR}}^{f+a}$ denote
          the CP-averaged branching ratios without and with the contributions
          from weak annihilation, respectively. The
          experimental data is from PDG2000; ${\ast}$ denotes
          uncorrelated averages of Table \ref{tab3}.}
 \label{tab4}
 \begin{tabular}{lcccccccccc}
   & \multicolumn{3}{c}{${\mu}=m_{b}/2$}
   & \multicolumn{3}{c}{${\mu}=m_{b}  $}
   & \multicolumn{3}{c}{${\mu}=2m_{b} $}
   & \\ \cline{2-4} \cline{5-7} \cline{8-10}
     \multicolumn{1}{c}{Decay}
   & NF & \multicolumn{2}{c}{QCDF}
   & NF & \multicolumn{2}{c}{QCDF}
   & NF & \multicolumn{2}{c}{QCDF}
   & \multicolumn{1}{c}{Exp.}
     \\ \cline{2-2} \cline{3-4}
        \cline{5-5} \cline{6-7}
        \cline{8-8} \cline{9-10}
     \multicolumn{1}{c}{Modes}
   & ${\rm BR}$ & ${{\rm BR}}^{f}$ & ${{\rm BR}}^{f+a}$
   & ${\rm BR}$ & ${{\rm BR}}^{f}$ & ${{\rm BR}}^{f+a}$
   & ${\rm BR}$ & ${{\rm BR}}^{f}$ & ${{\rm BR}}^{f+a}$
   & \\ \hline
   ${\overline{B}}^{0}{\to}{\pi}^{+}{\pi}^{-}$ &
       9.74  & 10.06 & 10.49 &
       9.09  & 9.67  & 9.97  &
       8.68  & 9.36  & 9.58  &
   ${4.4{\pm}0.9}^{\ast}$  \\
   $B^{-}{\to}{\pi}^{-}{\pi}^{0}$ &
       5.58 & 5.26 & --- &
       6.20 & 5.36 & --- &
       6.76 & 5.52 & --- &
   ${5.7{\pm}1.5}^{\ast}$ \\
   $B^{-}{\to}{\pi}^{-}{\eta}$    &
       3.41 & 3.27 & 3.22 &
       3.59 & 3.28 & 3.26 &
       3.74 & 3.34 & 3.33 & $<15 $ \\
   $B^{-}{\to}{\pi}^{-}{\eta}^{\prime}$   &
       2.26 & 2.23 & 2.18 &
       2.53 & 2.26 & 2.24 &
       2.71 & 2.35 & 2.34 & $<12$ \cite{belle} \\
   ${\overline{B}}^{0}{\to}{\pi}^{0}{\eta}$    &
       0.17 & 0.17 & 0.18 &
       0.12 & 0.15 & 0.16 &
       0.07 & 0.14 & 0.15 & $<8$  \\
   ${\overline{B}}^{0}{\to}{\pi}^{0}{\eta}^{\prime}$   &
       0.038 & 0.048 & 0.054 &
       0.037 & 0.045 & 0.053 &
       0.026 & 0.049 & 0.057 & $<11$ \\
   ${\overline{B}}^{0}{\to}{\pi}^{0}{\pi}^{0}$ &
       0.12 & 0.11  & 0.18  &
       0.15 & 0.099 & 0.13  &
       0.23 & 0.10  & 0.12  & $<9.3$ \\
   ${\overline{B}}^{0}{\to}{\eta} {\eta}$   &
       0.077 & 0.078 & 0.104 &
       0.095 & 0.076 & 0.093 &
       0.12  & 0.081 & 0.093 & $<18$ \\
   ${\overline{B}}^{0}{\to}{\eta}^{\prime}{\eta}^{\prime}$   &
       0.009 & 0.016 & 0.022 &
       0.029 & 0.015 & 0.022 &
       0.048 & 0.021 & 0.027 & $<47$ \\
   ${\overline{B}}^{0}{\to}{\eta} {\eta}^{\prime}$   &
       0.059 & 0.073 & 0.066 &
       0.11  & 0.073 & 0.073 &
       0.15  & 0.086 & 0.09  & $<27$ \\
   ${\overline{B}}^{0}{\to}K^{-}{\pi}^{+}$  &
       10.66 & 9.56 & 10.45 &
        6.40 & 8.59 & 9.13  &
         3.4 & 7.78 & 8.09  &
   ${17.4{\pm}1.5}^{\ast}$ \\
   ${\overline{B}}^{0}{\to}{\overline{K}}^{0}{\pi}^{0}$  &
        4.99 & 4.52 & 4.96 &
        2.90 & 4.01 & 4.27 &
        1.44 & 3.57 & 3.72 &
   ${10.3{\pm}2.6}^{\ast}$ \\
   $B^{-}{\to}K^{-}{\pi}^{0}$   &
        7.65 & 7.04 & 7.56 &
        4.83 & 6.36 & 6.67 &
        2.82 & 5.78 & 5.96 &
   ${12.1{\pm}1.6}^{\ast}$  \\
   $B^{-}{\to}{\overline{K}}^{0}{\pi}^{-}$  &
        14.43 & 13.45 & 14.48 &
          8.8 & 11.99 & 12.61 &
         4.79 & 10.72 & 11.08 &
   ${17.3{\pm}2.5}^{\ast}$  \\
   $B^{-}{\to}{K}^{-}{\eta}$    &
         2.91 & 2.67 & 2.78 &
         1.68 & 2.31 & 2.37 &
         0.94 & 1.96 & 2.0  & $<14$ \\
   $B^{-}{\to}{K}^{-}{\eta}^{\prime}$   &
         16.6  & 16.72 & 17.82 &
         12.47 & 15.74 & 16.87 &
          7.49 & 15.75 & 16.55 &
   $74.9{\pm}6.6^{\ast}$ \\
   ${\overline{B}}^{0}{\to}{\overline{K}}^{0}{\eta}$    &
          1.93 & 1.70 & 1.79 &
           1.0 & 1.45 & 1.50 &
          0.45 & 1.21 & 1.24 & $<33$ \\
   ${\overline{B}}^{0}{\to}{\overline{K}}^{0}{\eta}^{\prime}$   &
         17.07 & 17.33 & 18.13 &
         12.82 & 16.26 & 17.12 &
          7.79 & 16.14 & 16.73 &
   $59.6{\pm}9.8^{\ast}$ \\
   $B^{-}{\to}K^{-}K^{0}$     &
          0.78 & 0.75 & 0.79 &
          0.47 & 0.67 & 0.69 &
          0.26 & 0.59 & 0.61 & $<2.5$ \cite{babar} \\
   ${\overline{B}}^{0}{\to}{\overline{K}}^{0}K^{0}$   &
          0.73 & 0.71 & 0.84 &
          0.44 & 0.62 & 0.70 &
          0.24 & 0.56 & 0.60 & $<17$ \\
   ${\overline{B}}^{0}{\to}K^{+}K^{-}$     &
         --- & --- & 0.044 &
         --- & --- & 0.033 &
         --- & --- & 0.03  & $<2.5$ \cite{babar}
 \end{tabular}
 \end{table}

 \begin{table}
 \caption{Experimental data of CP-violating asymmetries ${\cal A}_{CP}$
          for $B{\to}{\pi}K,K{\eta}^{\prime}$,
          $a_{{\epsilon}^{\prime}}$,
          and $a_{{\epsilon}+{\epsilon}^{\prime}}$ for decay
          ${\overline{B}}^{0}{\to}{\pi}^{+}{\pi}^{-}$ in unit of percent.}
 \label{tab5}
 \begin{tabular}{rcccc}
  & \multicolumn{1}{c}{CLEO}
  & \multicolumn{1}{c}{BaBar}
  & \multicolumn{1}{c}{Belle} \\ \hline
 ${\cal A}_{CP}({\pi}^{\pm}K^{\mp})$
 & $-4{\pm}16$
 & $-7\pm 8 \pm 2$
 & $4.4^{+18.6+1.8}_{-16.7-2.1}$ \\
 ${\cal A}_{CP}({\pi}^{0}K^{\pm})$
 & $-29{\pm}23$
 & $0{\pm}18{\pm}4$
 & $-5.9^{+22.2+5.5}_{-19.6-1.7}$ \\
 ${\cal A}_{CP}({\pi}^{\pm}K^{0})$
 & $18{\pm}24$
 & $-21{\pm}18{\pm}3$
 & $9.8^{+43.0+2.0}_{-34.3-6.3}$ \\
 ${\cal A}_{CP}(K^{\pm}{\eta}^{\prime})$
 & $3{\pm}12$
 & & $6\pm 15 \pm 1$ \\ \hline
 $a_{{\epsilon}^{\prime}}({\pi}^{+}{\pi}^{-})$
 & & $-25^{+45}_{-47}{\pm}14$ & \\
 $a_{{\epsilon}+{\epsilon}^{\prime}}({\pi}^{+}{\pi}^{-})$
 & & $  3^{+53}_{-56}{\pm}11$ &
 \end{tabular}
 \end{table}

 \begin{table}
 \caption{The dependence of time-integrated CP-violating asymmetries
         $a_{{\epsilon}^{\prime}}$,
         $a_{{\epsilon}+{\epsilon}^{\prime}}$,
         and ${\cal A}_{CP}$ (in units of percent) for ${\overline{B}}^{0}
         (B^{0}){\to}{\pi}^{0}{\pi}^{0},{\pi}^{0}{\eta}^{({\prime})},
         {\eta}^{({\prime})}{\eta}^{({\prime})}$ on renormalization scale
         ${\mu}$, with the default values of various parameters in the BBNS
         approach.}
 \label{tab6}
 \begin{tabular}{lccccccc}
 \multicolumn{1}{c}{Decay Modes} & scale ${\mu}$ &
 $a_{{\epsilon}^{\prime}}^{f}$ &
 $a_{{\epsilon}^{\prime}}^{f+a}$ &
 $a_{{\epsilon}+{\epsilon}^{\prime}}^{f}$ &
 $a_{{\epsilon}+{\epsilon}^{\prime}}^{f+a}$ &
 ${\cal A}_{CP}^{f}$ &
 ${\cal A}_{CP}^{f+a}$  \\ \hline
    & $m_{b}/2$
    & -57.3 & -69.3 & -32.0 &  26.5 & -52.6 & -32.6 \\
 ${\overline{B}}^{0}(B^{0}){\to}{\pi}^{0}{\pi}^{0}$
    & $m_{b}$
    & -45.1 & -79.0 & -66.8 & -12.4 & -61.3 & -57.4 \\
    & $2m_{b}$
    & -38.3 & -82.4 & -89.2 & -46.3 & -67.5 & -75.8 \\ \hline
    & $m_{b}/2$
    & 35.5 & 62.7 & 9.72 & -30.7 & 27.8 & 26.3 \\
 ${\overline{B}}^{0}(B^{0}){\to}{\pi}^{0}{\eta}$
    & $m_{b}$
    & 31.5 & 63.3 & 12.2 & -19.3 & 26.4 & 32.1 \\
    & $2m_{b}$
    & 29.5 & 65.6 & 15.8 & -9.75 & 26.8 & 38.1 \\ \hline
    & $m_{b}/2$
    & 44.2 & 84.8 & 25.9 & -35.3 & 41.2 & 38.5 \\
 ${\overline{B}}^{0}(B^{0}){\to}{\pi}^{0}{\eta}^{\prime}$
    & $m_{b}$
    & 41.8 & 81.0 & 19.6 & -28.2 & 36.6 & 39.4 \\
    & $2m_{b}$
    & 37.1 & 79.3 & 18.3 & -17.4 & 33.0 & 43.5 \\ \hline
    & $m_{b}/2$
    & 55.5 & 47.3 & 58.8 & 51.4 & 64.2 & 55.4 \\
 ${\overline{B}}^{0}(B^{0}){\to}{\eta}{\eta}$
    & $m_{b}$
    & 43.1 & 36.4 & 71.1 & 69.9 & 61.9 & 57.0 \\
    & $2m_{b}$
    & 34.7 & 26.6 & 82.6 & 84.4 & 62.0 & 57.6 \\ \hline
    & $m_{b}/2$
    & 30.7 & 76.8 & 95.2 & 52.7 & 65.4 & 75.2 \\
 ${\overline{B}}^{0}(B^{0}){\to}{\eta}^{\prime}{\eta}^{\prime}$
    & $m_{b}$
    & 37.8 & 65.6 & 92.3 & 71.9 & 68.6 & 77.1 \\
    & $2m_{b}$
    & 36.7 & 48.1 & 92.9 & 87.0 & 68.2 & 72.8 \\ \hline
    & $m_{b}/2$
    & 55.6 & 49.5 & 74.1 & 77.8 & 71.5 & 69.3 \\
 ${\overline{B}}^{0}(B^{0}){\to}{\eta}{\eta}^{\prime}$
    & $m_{b}$
    & 46.0 & 35.3 & 81.7 & 87.3 & 68.9 & 64.6 \\
    & $2m_{b}$
    & 37.4 & 23.5 & 88.9 & 94.2 & 66.8 & 60.2
 \end{tabular}
 \end{table}

 \begin{table}
 \caption{CP-violating asymmetries $a_{{\epsilon}^{\prime}}({\%})$ and
          $a_{{\epsilon}+{\epsilon}^{\prime}}({\%})$ for neutral $B$
          meson decays with default values of various parameters using
          the BBNS approach.}
 \label{tab7}
  \begin{tabular}{lcccccccccccc}
     \multicolumn{1}{c}{Decay}
   & \multicolumn{4}{c}{${\gamma}=60^{\circ} $}
   & \multicolumn{4}{c}{${\gamma}=90^{\circ} $}
   & \multicolumn{4}{c}{${\gamma}=120^{\circ}$}
     \\ \cline{2-5} \cline{6-9} \cline{10-13}
     \multicolumn{1}{c}{Modes}
   & $a_{{\epsilon}^{\prime}}^{f}$
   & $a_{{\epsilon}^{\prime}}^{f+a}$
   & $a_{{\epsilon}+{\epsilon}^{\prime}}^{f}$
   & $a_{{\epsilon}+{\epsilon}^{\prime}}^{f+a}$
   & $a_{{\epsilon}^{\prime}}^{f}$
   & $a_{{\epsilon}^{\prime}}^{f+a}$
   & $a_{{\epsilon}+{\epsilon}^{\prime}}^{f}$
   & $a_{{\epsilon}+{\epsilon}^{\prime}}^{f+a}$
   & $a_{{\epsilon}^{\prime}}^{f}$
   & $a_{{\epsilon}^{\prime}}^{f+a}$
   & $a_{{\epsilon}+{\epsilon}^{\prime}}^{f}$
   & $a_{{\epsilon}+{\epsilon}^{\prime}}^{f+a}$ \\ \hline
   ${\pi}^{+}{\pi}^{-}$ &
       4.17  & 11.5  &  55.1  & 54.8 &
       5.85  & 16.1  & -30.7  &-30.6 &
       6.64  & 17.7  & -92.0  &-91.1 \\
   ${\pi}^{0}{\eta}$    &
       31.2  & 60.8  & 11.4  & -21.1 &
       25.0  & 43.5  & 7.27  & -22.1 &
       16.6  & 27.3  & 4.21  & -16.0 \\
   ${\pi}^{0}{\eta}^{\prime}$   &
       41.6  & 77.4  & 18.4  & -31.8 &
       34.7  & 54.5  & 11.5  & -35.8 &
       23.5  & 34.0  & 6.54  & -26.3 \\
   ${\pi}^{0}{\pi}^{0}$ &
       -41.5  & -76.6  & -65.3  & -16.0 &
       -25.6  & -56.9  & -49.8  & -23.7 &
       -15.1  & -36.4  & -32.1  & -18.8 \\
   ${\eta} {\eta}$   &
       45.7  & 38.5  & 70.6  & 69.9 &
       53.6  & 44.4  & 60.2  & 61.4 &
       47.2  & 38.4  & 43.3  & 45.0 \\
   ${\eta}^{\prime}{\eta}^{\prime}$ &
       41.7  & 70.3  & 89.7  & 68.4 &
       65.7  & 86.9  & 65.3  & 42.9 &
       89.9  & 81.1  & 38.4  & 21.0 \\
   ${\eta} {\eta}^{\prime}$  &
       49.7  & 38.4  & 81.0  & 87.4 &
       65.5  & 53.4  & 68.8  & 80.5 &
       66.1  & 58.0  & 50.7  & 65.7 \\
   $K^{0}_{s}{\pi}^{0}$      &
       1.78  & 2.37  & 77.0  & 76.6 &
       2.14  & 2.83  & 74.1  & 73.7 &
       1.92  & 2.54  & 58.8  & 58.3 \\
   ${K}^{0}_{s}{\eta}$      &
      3.08  & 3.62  & 77.8  & 77.4 &
      3.72  & 4.36  & 75.1  & 74.7 &
      3.38  & 3.95  & 59.9  & 59.3 \\
    ${K}^{0}_{s}{\eta}^{\prime}$     &
      -1.92  & -2.20  & 73.9  & 74.3 &
      -2.25  & -2.58  & 70.3  & 70.8 &
      -1.97  & -2.26  & 54.7  & 55.2 \\
   ${\overline{K}}^{0}K^{0}$   &
          21.0 & 20.8 & 9.08 & 3.74 &
          16.6 & 16.1 & 6.30 & 2.16 &
          10.9 & 10.5 & 3.86 & 1.17 \\
   $K^{+}K^{-}$  &
         --- & 0 & --- & 72.9 &
         --- & 0 & --- & 3.47 &
         --- & 0 & --- & -68.0
 \end{tabular}
 \end{table}

 \begin{table}
 \caption{CP-violating asymmetries ${\cal A}_{CP}({\%})$ for decays of
         $B{\to}PP$ with default values of various parameters using the BBNS
         approach.}
         \label{tab8}
 \begin{tabular}{lcccccc}
     \multicolumn{1}{c}{Decay}
   & \multicolumn{2}{c}{${\gamma}=60^{\circ} $}
   & \multicolumn{2}{c}{${\gamma}=90^{\circ} $}
   & \multicolumn{2}{c}{${\gamma}=120^{\circ}$}
     \\ \cline{2-3} \cline{4-5} \cline{6-7}
     \multicolumn{1}{c}{modes}
   & \multicolumn{1}{c}{${\cal A}_{CP}^{f}$}
   & \multicolumn{1}{c}{${\cal A}_{CP}^{f+a}$}
   & \multicolumn{1}{c}{${\cal A}_{CP}^{f}$}
   & \multicolumn{1}{c}{${\cal A}_{CP}^{f+a}$}
   & \multicolumn{1}{c}{${\cal A}_{CP}^{f}$}
   & \multicolumn{1}{c}{${\cal A}_{CP}^{f+a}$} \\ \hline
   ${\overline{B}}^{0}(B^{0}){\to}{\pi}^{+}{\pi}^{-}$
    & 28.9  & 33.6
    &-10.8  &-4.08
    &-39.6  &-31.9  \\
   $B^{\pm}{\to}{\pi}^{\pm}{\pi}^{0}$
    & 0.07  & ---
    & 0.09  & ---
    & 0.08  & ---  \\
   $B^{\pm}{\to}{\pi}^{\pm}{\eta}$
    & 11.5  & 25.5
    & 18.6  & 40.3
    & 27.2  & 55.4  \\
   $B^{\pm}{\to}{\pi}^{\pm}{\eta}^{\prime}$
    & 5.08  & 19.3
    & 7.55  & 28.1
    & 9.16  & 32.9 \\
   ${\overline{B}}^{0}(B^{0}){\to}{\pi}^{0}{\eta}$
    & 25.8  & 29.7
    & 19.8  & 17.9
    & 12.8  & 10.2  \\
   ${\overline{B}}^{0}(B^{0}){\to}{\pi}^{0}{\eta}^{\prime}$
    & 35.9  & 35.4
    & 28.1  & 18.5
    & 18.5  & 9.60  \\
   ${\overline{B}}^{0}(B^{0}){\to}{\pi}^{0}{\pi}^{0}$
    & 58.1  & 57.6
    & 40.4  & 48.4
    & 25.2  & 32.7  \\
   ${\overline{B}}^{0}(B^{0}){\to}{\eta} {\eta}$
    & 63.5  & 58.4
    & 63.7  & 58.2
    & 51.4  & 46.5  \\
    ${\overline{B}}^{0}(B^{0}){\to}{\eta}^{\prime}{\eta}^{\prime}$
    & 69.9  & 78.4
    & 74.0  & 77.1
    & 76.9  & 62.9  \\
   ${\overline{B}}^{0}(B^{0}){\to}{\eta} {\eta}^{\prime}$
    & 71.0  & 66.7
    & 75.5  & 73.2
    & 67.3  & 69.2  \\
   ${\overline{B}}^{0}(B^{0}){\to}K^{\mp}{\pi}^{\pm}$
    &-6.53  &-16.1
    &-5.91  &-14.8
    &-4.20  &-10.7  \\
   ${\overline{B}}^{0}(B^{0}){\to}K^{0}_{S}{\pi}^{0}$
    & 37.8  & 38.0
    & 36.7  & 36.9
    & 29.3  & 29.4 \\
   $B^{\mp}{\to}K^{\mp}{\pi}^{0}$
    &-7.24  &-15.0
    &-6.69  &-14.1
    &-4.83  &-10.3  \\
   $B^{\mp}{\to}K^{0}_{S}{\pi}^{\mp}$
    &-0.84  &-1.05
    &-0.99  &-1.24
    &-0.88  &-1.10  \\
   $B^{\mp}{\to}{K}^{\mp}{\eta}$
    & 8.35  & 14.0
    & 12.8  & 21.1
    & 16.3  & 26.3  \\
   $B^{\mp}{\to}{K}^{\mp}{\eta}^{\prime}$
    &-3.19  &-7.50
    &-3.41  &-8.12
    &-2.75  &-6.62 \\
   ${\overline{B}}^{0}(B^{0}){\to}K^{0}_{S}{\eta}$
    & 39.1  & 39.2
    & 38.2  & 38.4
    & 30.7  & 30.8  \\
   ${\overline{B}}^{0}(B^{0}){\to}K^{0}_{S}{\eta}^{\prime}$
    & 34.0  & 33.9
    & 32.0  & 32.0
    & 24.8  & 24.8  \\
   $B^{\mp}{\to}K^{\mp}K^{0}_{S}$
    & 21.0  & 26.4
    & 16.6  & 20.2
    & 10.9  & 13.1  \\
   ${\overline{B}}^{0}(B^{0}){\to}{\overline{K}}^{0}K^{0}$
    & 18.0  & 15.4
    & 13.8  & 11.5
    & 8.97  & 7.38  \\
   ${\overline{B}}^{0}(B^{0}){\to}K^{+}K^{-}$
    & ---  &  34.7
    & ---  &  1.65
    & ---  & -32.4
 \end{tabular}
 \end{table}

 \begin{table}
 \caption{The dependence of CP-averaged branching ratio and CP asymmetries
          for $B^{\pm}{\to}K^{\pm}{\pi}^{0}$ on ${\varrho}$, ${\phi}$ in
          endpoint divergence ${\int}^{1}_{0}~dx/x$. ${{\rm BR}}^{f}$ and
          ${{\rm BR}}^{f+a}$ have the same meanings as those in Table
          \ref{tab4}. ${\cal A}_{CP}^{f}$ and ${\cal A}_{CP}^{f+a}$ can be
          defined in a similar way. The results are calculated with
          default values of other parameters at ${\mu}=m_{b}$ using the BBNS
          approach.}
 \label{tab9}
 \begin{tabular}{rccccccccc}
  \multicolumn{1}{c}{${\varrho}$}
    & 0 & \multicolumn{4}{c}{3} & \multicolumn{4}{c}{6}
    \\ \cline{2-2} \cline{3-6} \cline{7-10}
  \multicolumn{1}{c}{${\phi}(\text{deg})$}
    & $0{\sim}360$ & 0 & 90 & 180 & 270 & 0 & 90 & 180 & 270 \\ \hline
  ${{\rm BR}}^{f}(10^{-6})  $ & 6.37 & 6.31 & 6.36 & 6.43 & 6.38
                               & 6.25 & 6.35 & 6.49 & 6.39 \\
  ${{\rm BR}}^{f+a}(10^{-6})$ & 7.51 & 11.8 & 6.75 & 6.43 & 6.18
                               & 21.3 & 4.63 & 7.89 & 3.48 \\
  ${\cal A}_{CP}^{f}({\%})   $ & 8.28 & 8.45 & 7.15 & 8.11 & 9.40
                               & 8.63 & 6.02 & 7.94 & 10.5 \\
  ${\cal A}_{CP}^{f+a}({\%}) $ & 7.24 & 5.15 & 14.2 & 8.11 & 1.47
                               & 3.32 & 29.5 & 6.81 &-12.4
 \end{tabular}
 \end{table}

\newpage

 \begin{figure}
 \begin{center}
 \begin{picture}(300,200)
 \put(-60,-20){\epsfxsize150mm\epsfbox{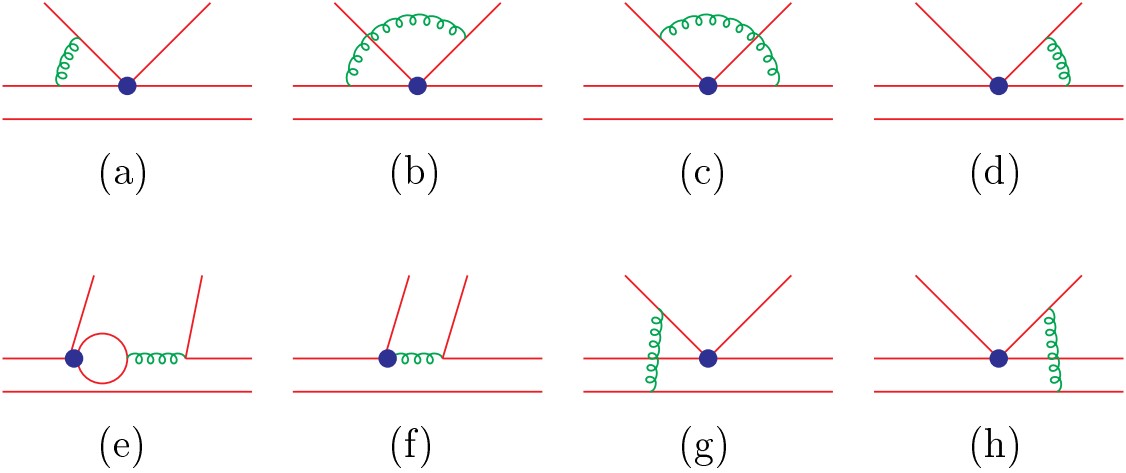}}
 \end{picture}
 \end{center}
 \vspace{2cm}
 \caption{Order of ${\alpha}_{s}$ corrections to hard-scattering kernels.
 The upward quark lines represent the ejected light meson
 from $b$ quark weak decays. These diagrams are commonly called vertex
 corrections, penguin corrections and hard spectator diagrams for
 Fig.(a)-(d), (e)-(f) and (g)-(h), respectively.}
 \end{figure}

\vspace{2cm}
 \begin{figure}
 \begin{center}
 \begin{picture}(300,120)
 \put(-60,0){\epsfxsize150mm\epsfbox{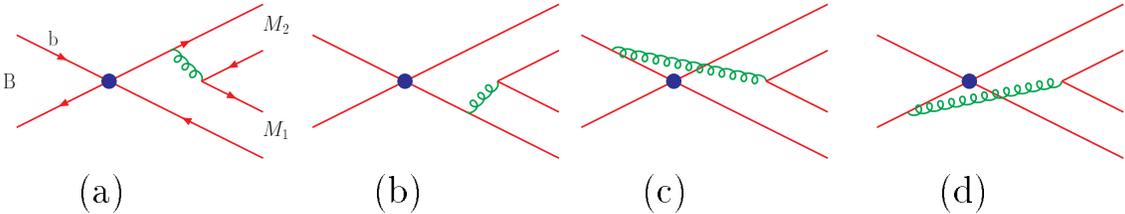}}
 \end{picture}
 \end{center}
 \vspace{1cm}
 \caption{Order of ${\alpha}_{s}$ corrections to annihilation diagrams
 for $B{\rightarrow}PP$ }
 \end{figure}

\newpage
\begin{figure}
 \begin{center}
 \begin{picture}(300,490)
 \put(-90,100){\epsfxsize 160 mm\epsfbox{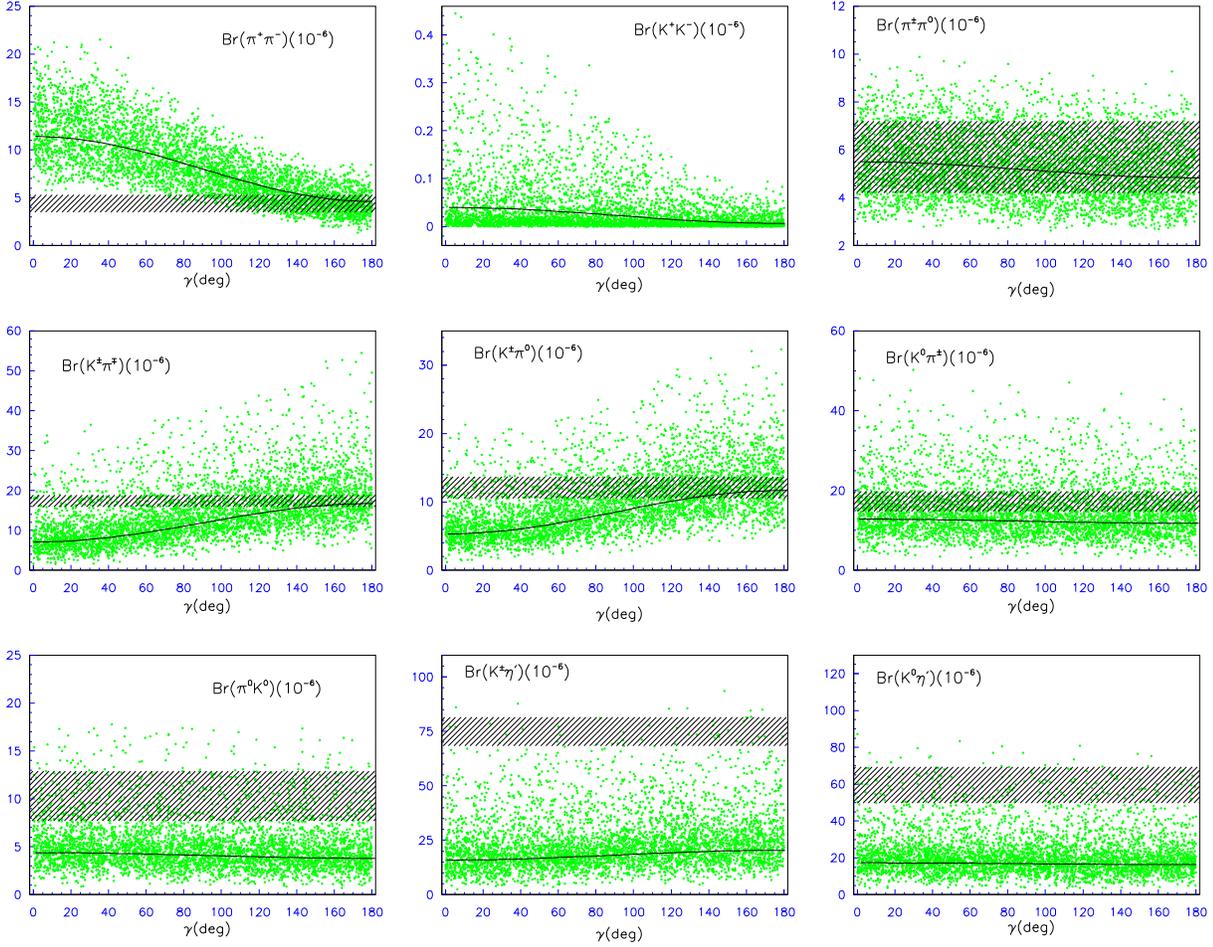}}
 \end{picture}
 \end{center}
 \vspace{-1cm}
 \caption{CP-averaged branching ratios as functions of ${\gamma}$. The
          solid lines are drawn with the default values at the scale of
          ${\mu}=m_{b}$, the horizontal slashed-line bands correspond to
          experimental data within one standard error, and the dot-shades
          denote the variation of the theory input parameters, including
          the CKM elements, the form factors, the uncertainty from weak
          annihilation and hard spectator scattering.}
 \end{figure}

\newpage
 \begin{figure}
 \begin{center}
 \begin{picture}(300,280)(0,40)
 \put(-85,-180){\epsfxsize 160mm\epsfbox{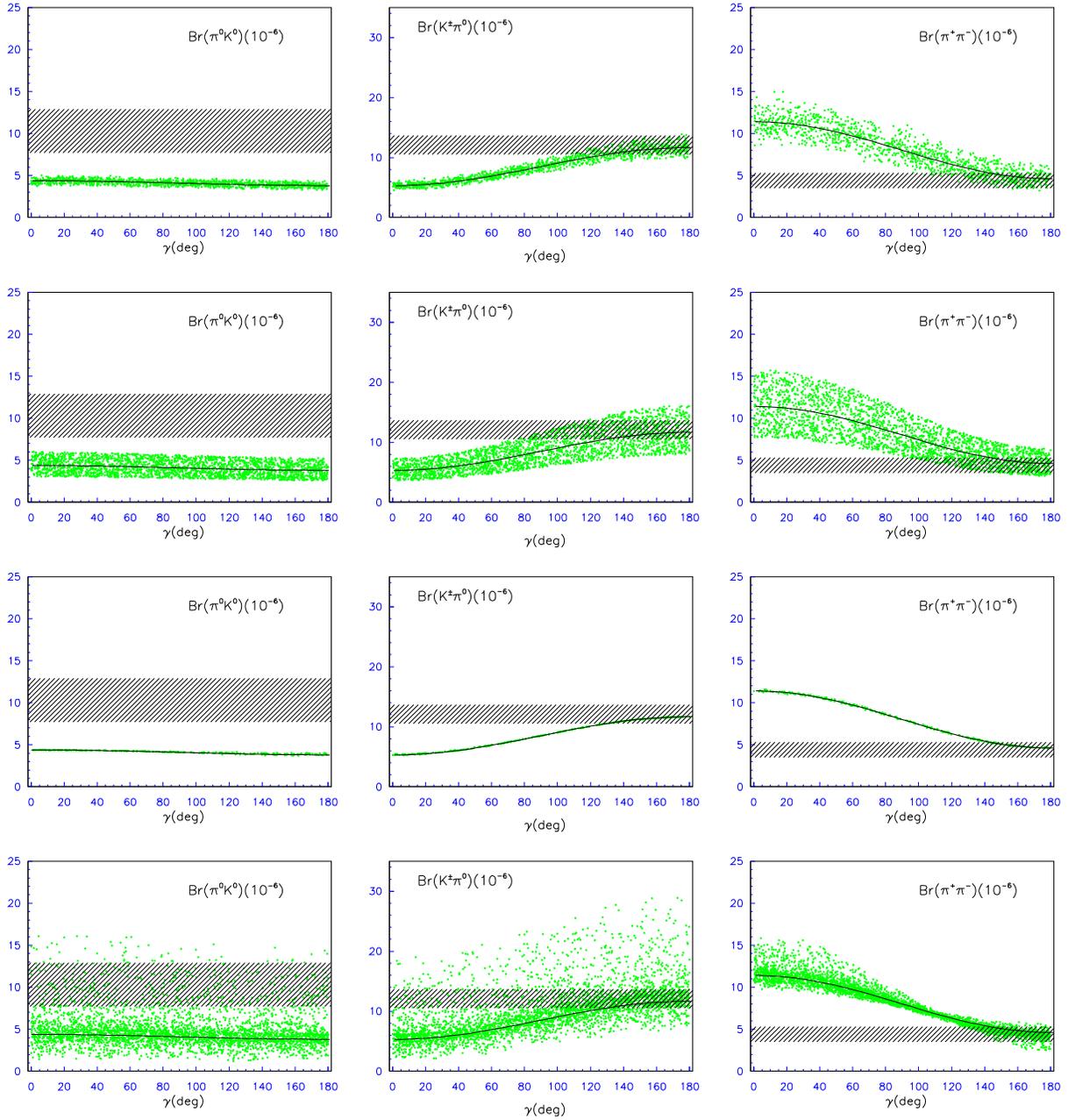}}
 \end{picture}
 \end{center}
 \vspace{8cm}
 \caption{The dependence of CP-averaged branching ratios for
 $B^0\to \pi^0 K^0$, $B^{\pm}{\to}K^{\pm}{\pi}^{0}$ and $B^0 \to \pi^+
 \pi^-$ on the variations of input parameters under the QCDF approach
 at the renormalization scale ${\mu}=m_{b}$. The lines and bands have
 the same meaning as in Fig. 3, and the dot-shades correspond to
 the variation of  the CKM matrix elements, in particular $\vert V_{ub} \vert$
  (the first row), form factors (the second row), endpoint divergences in
  the hard spectator scattering $X_H$(the third row) and weak annihilations $X_A$(the forth row).}
 \end{figure}

\newpage
  \begin{figure}
  \begin{center}
  \begin{picture}(300,550)
  \put(-90,-40){\epsfxsize 160mm\epsfbox{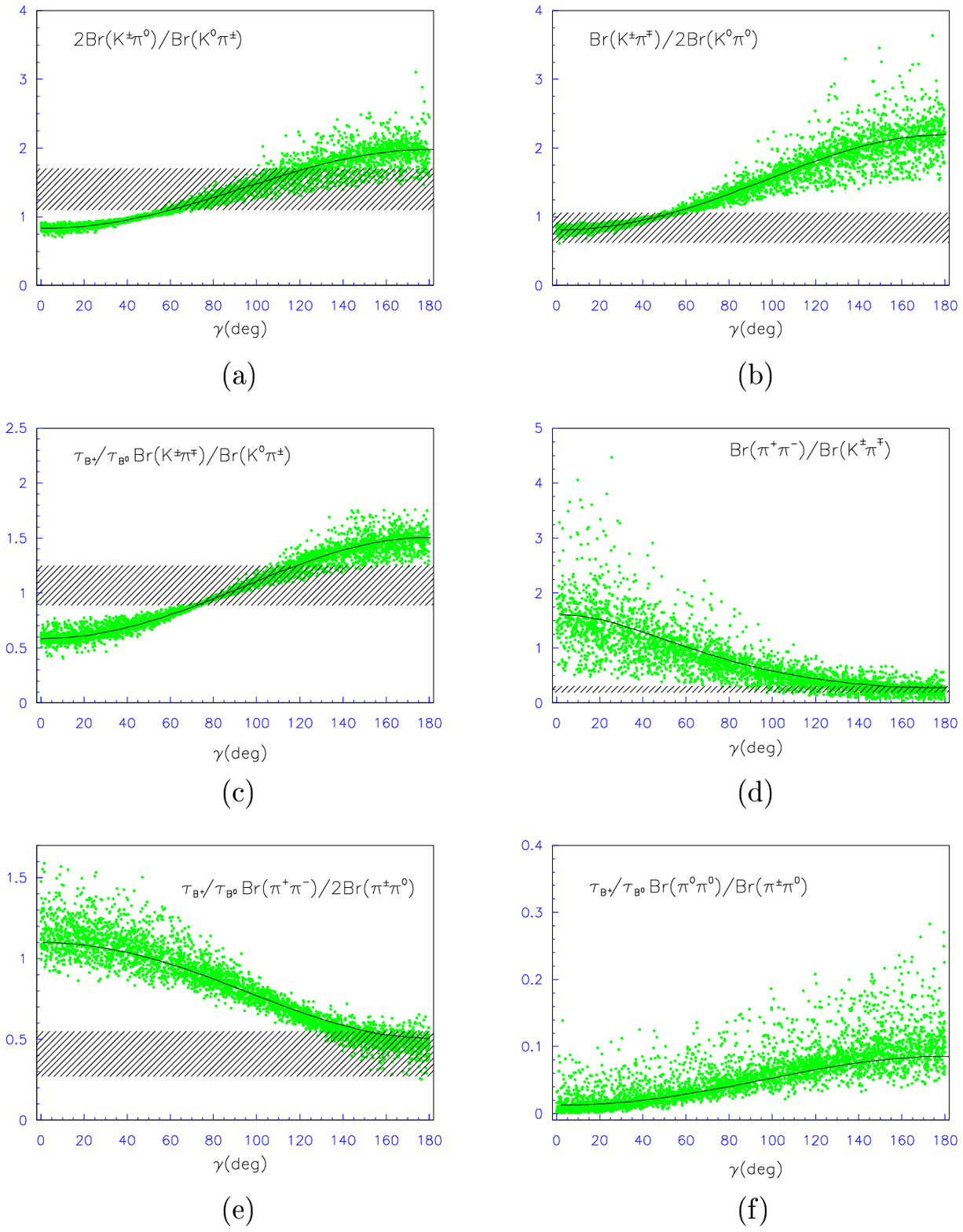}}
  \end{picture}
  \end{center}
  \vspace{1cm}
  \caption{Ratios of CP-averaged branching fractions versus ${\gamma}$.
           The lines, bands and dot-shades have the same meaning as in
           Fig. 3.}
 \end{figure}


\begin{thebibliography}{99}
 \bibitem{cleo} CLEO Collaboration, D. Cronin-Hennessy {\it et al.},
         Phys. Rev. Lett. {\bf 85} 515 (2000); CLEO Collaboration,
         S. Chen {\it et al.}, {\it iibid.} {\bf 85} 525 (2000);
         CLEO Collaboration, S.J. Richichi {\it et al.},
         {\it ibid.} {\bf 85} 520 (2000).

 \bibitem{babar}Babar Collaboration, G. Cavoto, hep-ex/0105018;
               Babar Collaboration, B. Aubert {\it et al.},
           Phys. Rev. Lett.{\bf 87}, 151802 (2001); hep-ex/0107074, hep-ex/0108017;
              Babar Collaboration, J. Olsen, hep-ex/0011031.

 \bibitem{belle} Belle Collaboration, T. Iijima, hep-ex/0105005; Belle Collaboration, K. Abe
{\it et al.}, Phys. Rev. Lett. {\bf 87}, 101801, (2001); Phys.
Rev. {\bf D64}, 071101, (2001); Phys. Lett. {\bf B517}, 309
(2001).

 \bibitem{buras1}For a review, see G. Buchalla, A.J. Buras and
          M.E. Lautenbacher,
          Rev. Mod. Phys. {\bf 68} 1125 (1996) ;
          or A.J. Buras, hep-ph/9806471.

 \bibitem{BSW}
          M. Bauer and B. Stech,
          Phys. Lett. {\bf B152} 380 (1985);
          M.Bauer, B.Stech and M.Wirbel,
          Z. Phys. {\bf C34} (1987) 103.

\bibitem{bjorken} J.D. Bjorken, Nucl. Phys. B(Proc. Suppl.) {\bf
11}, 325 (1989).

\bibitem{GF}H.Y. Cheng,
          Phys .Lett. {\bf B335} 428 (1994);
         {\bf B395} 345 (1997);
          H.Y. Cheng and B. Tseng,
          Phys. Rev. {\bf D58} 094005 (1998);
         A. Ali and C. Greub,
         {\it ibid.} {\bf 57} 2996 (1998);
         A. Ali, J. Chay, C. Greub and P. Ko,
         Phys. Lett. {\bf B424} 161 (1998);
         M. Neubert, Nucl. Phys. B (Proc.Suppl.) {\bf 64}
         474 (1998);
         J. M. Soares, Phys. Rev. {\bf D51} 3518 (1995).

        \bibitem{lattice}
        APE Collaboration, A. Abada {\it et al.}, Phys. Lett. {\bf B365},
        275 (1996); J.M. Flynn and C.T. Sachrajda, {\it Heavy flavours
        II}, pp. 402-452, (hep-lat/9710057); UKQCD Collaboration,
        J.M. Flynn {\it et al.},  Nucl.
        Phys. $\bf B461$, 327 (1996); UKQCD Collaboration, L. Del Debbio {\it et al.},
        Phys. Lett.  {\bf B416}, 392 (1998).

       \bibitem{pball}
       P. Ball and V.M. Braun, Phys. Rev. {\bf D55}, 5561, (1997);
       J. High Energy Phys. {\bf 09}, 005, (1998).

      \bibitem{ruckle}
      A. Khodjamirian, R. R\"uckl, S. Weinzierl and O. Yakovlev, Phys. Lett.
      {\bf B410}, 275 (1997);
      A. Khodjamirian, R. R\"uckl and C.W. Winhart, Phys. Rev. {\bf D58}, 054013
      (1998); A. Khodjamirian, R. R\"uckl,
      {\it Heavy flavours II}, pp. 345-401,
      hep-ph/9801443;  A. Khodjamirian, R. R\"uckl, S. Weinzierl,
       C.W. Winhart, O. Yakovlev, Phys. Rev. {\bf D62}, 114002
       (2000)


 \bibitem{ali}
 A. Ali, G. Kramer and C.D. L\"{u},
         Phys. Rev. {\bf D58} 094009 (1998);
       {\bf 59} 014005 (1999).

\bibitem{chy}
 Y.H. Chen, H.Y. Cheng, B. Tseng and K.C. Yang, Phys. Rev. {\bf D60}, 094014 (1999).

 \bibitem{beneke}
       M. Beneke, G. Buchalla, M. Neubert and C.T. Sachrajda,
       Phys. Rev. Lett. {\bf 83}, 1914 (1999).

  \bibitem{osaka}
  M. Beneke, G. Buchalla, M. Neubert and C.T. Sachrajda,
  hep-ph/0007256.

\bibitem{ours2}
 D.S. Du, D.S. Yang and G.H. Zhu, hep-ph/0008216; Phys.
 Lett. {\bf B509}, 263 (2001); Phys. Rev. {\bf D64}, 014036
 (2001).

  \bibitem{0104110}
  M. Beneke, G. Buchalla, M. Neubert and C.T. Sachrajda, Nucl. Phys. {\bf B606},245 (2001).

       \bibitem{chay}
       J. Chay, Phys.Lett. {\bf B476}, 339 (2000).

        \bibitem{nucl}
       M. Beneke, G. Buchalla, M. Neubert and C.T. Sachrajda, Nucl. Phys.
       {\bf B591}, 313 (2000).

       \bibitem{our}
        D.S. Du, D.S. Yang and G.H. Zhu, Phys. Lett. {\bf B488}, 46
        (2000).

        \bibitem{ymz}
        T. Muta, A. Sugamoto, M.Z. Yang and Y.D. Yang, Phys. Rev. {\bf D62}, 094020
        (2000).


       \bibitem{ymz1}
        M.Z. Yang and Y.D. Yang, Phys. Rev. {\bf D62}, 114019
        (2000); Nucl. Phys. {\bf B609}, 469 (2001).

      \bibitem{hxg}
     X.G. He, J.P. Ma and C.Y. Wu, Phys. Rev. {\bf D63}, 094004 (2001).

       \bibitem{chay1}
       J. Chay, C. Kim, hep-ph/0009244.

           \bibitem{chy1}
       H.Y. Cheng and K.C. Yang, Phys. Rev. {\bf D63}, 074011,
       (2001); {\bf 64}, 074004 (2001).

      \bibitem{lihn}
 Y.Y. Keum, H.N. Li, A.I. Sanda, Phys. Rev. {\bf D63}, 054008
 (2001); Phys. Lett. {\bf B504}, 6 (2001).

  \bibitem{Sachrajda}
       S. Descotes-Genon and C.T. Sachrajda, hep-ph/0109260.

     \bibitem{BSS}
     M. Bander, D. Silverman and A. Soni, Phys. Rev. Lett. {\bf 43}, 242
(1979).


       \bibitem{t3}
       V.L. Chernyak and A.R. Zhitnitsky, Phys. Rep. {\bf 112}, 173
(1983);
        V.M. Braun and I.B. Filyanov, Z. Phys.{\bf C48}, 239 (1990);
        P. Ball, J. High Energy Phys. {\bf 01}, 010 (1999).

           \bibitem{beneke1}
       M. Beneke, J. Phys. {\bf G27}, 1069 (2001).

\bibitem{weizt}
  D.S. Du, C.S. Huang, Z.T. Wei and M.Z. Yang, Phys. Lett. {\bf B520}, 50 (2001).

  \bibitem{threshold}
  H.N. Li, hep-ph/0103305; T. Kurimoto, H.N. Li and A. I. Sanda,
  Phys. Rev. {\bf D65}, 014007 (2002).


       \bibitem{lucd}
       C.D. L\"u, K. Ukai and M.Z. Yang, Phys.Rev. {\bf D63}, 074009
       (2001).

 \bibitem{roma}
 M.Ciuchini {\it et al.}, J. High Energy Phys. {\bf 07}, 13 (2001).

\bibitem{feldman}
T. Feldmann, P. Kroll, B. Stech, Phys. Rev. {\bf D58}, 114006
(1998).

\bibitem{0010175}
P. Colangelo and A. Khodjamirian, in {\it At the Frontier of
Particle Physics / Handbook of QCD}, Boris Ioffe Festschrift,
edited by M. Shifman (World Scientific, Singapore, in press),
hep-ph/0010175.

\bibitem{lp01}
M. Neubert, talk given in the XXth International Symposium on
Lepton and Photon Interactions at High Energies, Roma, Italy,
2001, hep-ph/0110301.

\bibitem{0110093}
M. Neubert, talk presented at the International Workshop on QCD:
Theory and Experiment, Martina Franca, Italy, 2001, and at the 9th
International Symposium on Heavy Flavour Physics, Pasadena, CA,
2001, hep-ph/0110093.


 \bibitem{yyd}D.S. Du, C.S. Kim and Y.D. Yang,
       Phys. Lett. {\bf B426}, 133 (1998);
          A.L. Kagan and A.A. Petrov, hep-ph/9707354;
          A.L. Kagan, hep-ph/9806266;
          M.R. Ahmady, E. Kou and A. Sugamoto,
         Phys. Rev. {\bf D58}, 014015 (1998).


\bibitem{ff}
D.S. Du, D.S. Yang and G.H. Zhu, HEP \& NP, {\bf 26(1)}, 1-7
(2002)(in Chinese), hep-ph/9912201.

\bibitem{lihn2}
C.H. Chen, H.N. Li, Phys. Rev. {\bf D63}, 014003 (2001).


 \bibitem{fsi}
  W.S. Hou and K.C. Yang, Phys. Rev. Lett. {\bf 84}, 4806 (2000).


\bibitem{xing}
Z.Z. Xing, Phys. Lett. {\bf B493}, 301 (2000).


\bibitem{charming}
C. Isola {\it et al.}, Phys. Rev. {\bf D64}, 014029 (2001);
hep-ph/0110411.

 \bibitem{neubert}
 M. Neubert and J.R. Rosner, Phys. Lett. {\bf B441}, 403 (1998).

 \bibitem{fleisher}
 R. Fleischer and T. Mannel, Phys. Rev. {\bf D57}, 2752 (1998);
 A.J. Buras and R. Fleischer, Eur. Phys. J. {\bf C16}, 97 (2000).

 \bibitem{neubert1}
 M. Neubert, Phys. Lett. {\bf B424}, 152 (1998);

  \bibitem{fleisher1}
  R. Fleischer, Eur. Phys. J {\bf C16}, 87 (2000); hep-ph/0011323.

 \bibitem{bigi}
 I.I. Bigi and A.I. Sanda, Nucl. Phys. {\bf B281}, 41 (1987);
 Y. Nir and H.R. Quinn, Ann. Rev. Nucl. Part. Sci. {\bf 42}, 211 (1992);
           P.Ball, hep-ph/0010024.
 \end{thebibliography}
 \end{document}